\documentclass[aps,prb,reprint,superscriptaddress]{revtex4-2}

\usepackage{tikz}
\usepackage{scalerel}
\usepackage{amssymb}
\usepackage{suffix}
\usepackage{float}
\usepackage{mathtools}
\usepackage[utf8]{inputenc}
\usepackage{booktabs}
\usepackage{cases}
\usepackage[multiple]{footmisc}
\usepackage{dcolumn}
\usepackage{color,soul}
\usepackage{rotating}
\usepackage{perpage}
\usepackage{xcolor}
\usepackage{comment}
\usepackage{soul}
\usepackage{tikz}
\usepackage[T1]{fontenc}
\usepackage{etoolbox}
\usepackage{graphics}
\usepackage{siunitx}
\usepackage{hyperref}
\hypersetup{colorlinks}
\usepackage{float}	
\usepackage{collref}
\usepackage{multirow}
\usepackage{mathtools}
\usepackage{bm}
\usepackage{url}
\usepackage{wasysym}

\begin{document}
\title{Phases and phase transition in Grover's algorithm
with systematic noise}
\author{Sasanka Dowarah}
 \email{sasanka.dowarah@utdallas.edu}
 \affiliation{Department of Physics, The University of Texas at Dallas, Richardson, Texas 75080, USA}
\author{Chuanwei Zhang}
 \affiliation{Department of Physics, The University of Texas at Dallas, Richardson, Texas 75080, USA}
 \affiliation{Department of Physics, Washington University, St. Louis, Missouri, 63130, USA}
\author{Vedika Khemani}
\affiliation{Department of Physics, Stanford University, Stanford, California 94305, USA
}
\author{Michael Kolodrubetz}
\email{mkolodru@utdallas.edu}
 \affiliation{Department of Physics, The University of Texas at Dallas, Richardson, Texas 75080, USA}
\date{\today}
\begin{abstract}
While limitations on quantum computation under Markovian environmental noise are well-understood in generality, their behavior for different quantum circuits and noise realizations can be less universal. Here we consider a canonical quantum algorithm - Grover's algorithm for unordered search on $L$ qubits - in the presence of systematic noise. This allows us to write the evolution as a random Floquet unitary, which we show is well-characterized by random matrix theory (RMT). The RMT analysis enables analytical predictions for different regimes and critical disorders of the many-body dynamics. We find two critical disorders. At small disorder, finite-dimensional manifold remains nonergodic as long as the noise is smaller than an ergodicity breaking transition, which scales as $O(L^{-1})$. Computational power is lost at a much smaller disorder that scales as $O(2^{-L/2})$. We comment on relevance to non-systematic noise in realistic quantum computers, including cold atom, trapped ion, and superconducting platforms.
\end{abstract}
\maketitle
{\allowdisplaybreaks}
 \parskip 0 pt

\section{Introduction}
Significant progress has been made in the last few years toward building
noisy intermediate scale quantum (NISQ) computers \cite{Preskill2018quantumcomputingin}. These quantum devices have fewer qubits ranging from dozens to a few hundred and despite their limitations, they have been used to tackle certain classically challenging
 problems and emulate numerous many-body phases of matter
\cite{Fauseweh2024, RevModPhys.94.015004}. However, more conventional quantum algorithms designed for ideal quantum computers do not in general perform meaningful computations when run on a NISQ device, suggesting that general usage will require quantum error correction.

Until such a time as quantum error correction is achieved, it remains important to understand the role of noise in creating issues for known quantum algorithms. One such algorithm is Grover's algorithm for unstructured database search \cite{10.1145/237814.237866}. It has been the subject of intense study and has, in fact, been implemented on various NISQ devices \cite{Figgatt2017Complete3G,Zhang2021ImplementationOE,Zhang2022QuantumSO}. The effects of noise on Grover's search have been studied using different models
and efforts have been made towards putting limits on the performance of the algorithm. For example, Long \emph{et al}.~\cite{PhysRevA.61.042305} determined that the size of the database is constrained by $\mathcal{O}(1/\delta^{2})$ where $\delta$ is an error parameter representing gate imperfection. In the presence of random phase error in the oracle gate, Shenvi \emph{et al}.~\cite{PhysRevA.68.052313} have shown that in order to maintain the probability of success, the error must decrease as $k^{1/4}$ when the database size is increased by a factor of $k$.  In the case of depolarizing noise channel, Salas \cite{Salas2008NoiseEO} found that the noise threshold behaves as $N^{-1.1}$, where $N$ is the size of the database. Norman and Ataba \cite{PhysRevA.61.012301} determined that the allowed noise rate scales as $N^{-2/3}$ if Gaussian noise is added at each step of the algorithm. Finally, Shapira \emph{et al}.~\cite{PhysRevA.67.042301} showed that for Grover's algorithm with $L$ qubits, with unitary noise characterized by standard deviation $\epsilon$, must satisfy $\epsilon < \mathcal{O}(L^{-1/2}N^{-1/4})$ where $N = 2^{L}$, in order for the algorithm to maintain a significant efficiency. It was also shown that any NISQ algorithm will fail to achieve the quadratic speedup over classical algorithms in the unstructured database search problem\cite{Chen2023}.

These cases primarily studied unphysical noise models on the long-range many-qubit Grover gates and consistently found that exponentially small error rates prevent the algorithm from succeeding. These error models make generalizing the results rather challenging, leaving open the question of whether other physically realistic error models might have improved performance bounds. Furthermore, the standard Grover algorithm involves a call to an oracle which is a non-$k$-local operation that is unphysical in most realizations. In practice, such non-$k-$local gates are implemented by decomposing it in terms of $k-$local gates ($k=1, 2$ in most cases that act on one qubit, or two qubits that are not necessarily adjacent to each other). We will consider the impact of errors on a circuit in which the oracle is unraveled into single qubit gates and $2$-local unitary gates (cf.~Fig \ref{fig:Linear depth decomposition for L = 4.}), which comes at the cost of larger depth.

In order to get a better analytical handle on the error-tolerance of Grover's algorithm, we consider here the effects of \emph{systematic gate noise}, i.e., undesired noise in the quantum gates that is repeated systematically throughout the course of the algorithm. Systematic noise naturally occurs in most experiments due to calibration issues. For our purposes, it imbues the noisy algorithm with a time-periodic (Floquet) structure, such that we can find the exact scaling of target state dynamics, noise thresholds, and other physically relevant quantities by employing both analytical and numerical approaches. We show that these signatures can be captured by a simple random-matrix Hamiltonian whose perturbative effect on the Floquet spectrum describes gap-closing transitions that determine the response to noise. We find that for a database of size $N = 2^{L}$, the algorithm loses its computational power at a noise strength that scales as $\mathcal{O}(N^{-1/2})$. Additionally, an ergodicity transition occurs at a noise strength that scales as 
$\mathcal{O}(L^{-1})$ with the system size. We show that these analytical predictions align perfectly with the exact numerical solution and comment on the relevance to using Grover's algorithm -- as well as related amplitude amplification techniques -- in the NISQ era.
\section{Model}

We start by reviewing the basic formulation of Grover's algorithm. Suppose we want to search for an item $w$ in an unstructured database with $N=2^{L}$ items. For simplicity, let us assume that the target item $w$ is unique.
The search problem can be expressed by using an indicator function $f(u)$ that takes integer inputs $u$, with $0\leq u \leq N-1$, such that
\begin{equation}
f(u) = 
\begin{cases}
    1  & \text{if $u$ is the solution, i.e., $u=w$} \\
  0 & \text{otherwise.}
\end{cases}  \nonumber  
\end{equation}
In the case of Grover's algorithm, this indicator function is implemented using a unitary operator $U_w$, known as the oracle, whose action on the computational basis states is
\begin{equation}
    |u\rangle \xrightarrow{U_w} (-1)^{f(u)}|u\rangle.
\end{equation}
The complexity of an algorithm can be measured by the number of times the function $f(u)$ is evaluated. Classical algorithms require $\sim N/2$ ``time steps'' -- calls to a circuit that evaluates the function $f(u)$ -- on average to find the target state.  Grover's algorithm requires $\mathcal{O}(\sqrt{N})$ evaluations of the oracle $U_{w}$ to obtain the target state, and thus provides quadratic speed up of query complexity over the classical counterparts. 

Instead of the elements of the database directly, Grover's algorithm focuses on the indices of the elements, such that the database can be represented using $L$ qubits as $\{|0\rangle,|1\rangle,\cdots,|2^{L}-1\rangle \}$. The algorithm starts with the uniform superposition state $|x\rangle = \frac{1}{\sqrt{N}} \sum^{N-1}_{j=0} |j\rangle$, and then amplifies the amplitude of the target state $|w\rangle$ using the action of the two unitary operators. First, it applies the oracle, which flips the sign of all the states except for the state $|w\rangle$,
\begin{equation}
    U_{w}|k\rangle = (2|w\rangle \langle w|-I)|k \rangle =  \begin{cases}
                |k \rangle  & \text{if } k = w \\
                -|k \rangle   & \text{if } k \neq w
\end{cases}.
\end{equation}
Second is the ``diffusion'' operator $U_{x}$, which flips the amplitude of all the states about their mean,
\begin{eqnarray}
    U_{x}\sum_{k}\alpha_{k}|k\rangle &=& (2|x\rangle\langle x|-I)\sum_{k}\alpha_{k}|k\rangle,\nonumber\\
    &=& \sum_{k} (-\alpha_{k}+\sum_{j}\alpha_{j}/N)|k\rangle.
\end{eqnarray}
The combined action of these two reflections is written in terms of the Grover operator as $G_{w} = U_{x}U_{w}$, which gradually rotates the initial state $|x\rangle$ towards the target state $|w\rangle$. The action of $G_{w}$ on the state of the system can be interpreted geometrically by writing the state of the system at $t=0$ as
\begin{align}
|\psi(t=0)\rangle  = |x\rangle & = \cos\theta|\Bar{x}(w)\rangle+\sin\theta|w\rangle,\\
 \mbox{where}\hspace{5mm}|\Bar{x}(w)\rangle & = \frac{1}{\sqrt{N-1}}\sum_{j\neq w} |j\rangle,   
\end{align}
and $\sin\theta = 1/\sqrt{N}$ . Then the state of the system at time $t$ can be written as
\begin{eqnarray}
    |\psi(t)\rangle &=& (G_{w})^{t} |\psi(t=0)\rangle,\nonumber\\
    &&= \cos[(2t+1)\theta]|\Bar{x}(w)\rangle + \sin[(2t+1)\theta]|w\rangle.\label{eq:system_at_time_t}
\end{eqnarray}
In other words, $G_w$ rotates the wave function in the two-dimensional subspace spanned by $|\Bar{x}_{w}\rangle$ and $|w\rangle$ by angle $2\theta$. From Eq. \eqref{eq:system_at_time_t}, we get the probability of measuring the target $|w\rangle$ after $t$ steps as,
\begin{equation}
    P_{w}(t) = \sin^{2}[(2t+1)\theta].\label{eq:probability-target-no-noise}
\end{equation}
Therefore, to obtain the target state with maximum probability, one performs a measurement on the system at $t = \lfloor \frac{\pi}{4}\sqrt{N}\rfloor$ steps \cite{Boyer1996}. Not only does this scale faster than a classical computer; it has in fact been shown that Grover's algorithm is asymptotically optimal on an ideal quantum computer \cite{PhysRevA.60.2746}.

\subsection{Floquet spectrum}

Since Grover's algorithm involves the repeated application of an identical unitary $G_w$, it can be achieved by an appropriately chosen time-periodic Hamiltonian $H(t+T) = H(t)$. In other words, Grover's algorithm is an example of a Floquet problem \cite{OkaReview2019}.
This allows us to define the time-independent Floquet Hamiltonian $H_{F}$ as \cite{doi:10.1080/00018732.2015.1055918,FloquetSurL}
\begin{equation}
    G_{w} = \mathcal{T} e^{-i \int^{T}_{0} dt H(t)} = e^{-iH_{F}T}, \label{Floquet-Hamiltonian}
\end{equation}
where we choose units with $\hbar=1$ throughout. It follows that eigenstates $|u_\alpha\rangle$ of $G_w$ satisfy
\begin{equation}
     G_{w}|u_{\alpha}\rangle= e^{-i\phi_{\alpha}T}|u_{\alpha}\rangle.\label{eq:Grover-quasienergies-definition}
\end{equation}
The quasi-energies $\phi_{\alpha}$ are defined modulo $2\pi/T$, where period $T\equiv 1$ represents the time required to apply $G_{w}$ on the qubits. The actual value of time $T$ is irrelevant to this analysis because we can express everything using dimensionless phases rather than working with energies.

Before we write the exact eigenstates and quasi-energies for the Grover operator, we state a useful theorem that simplifies our analysis: the spectrum of the Grover operator is independent of the target state $w$. To prove this, consider conjugation by the unitary  matrix
\begin{equation}
    V_{w} = \prod_{j \in w} X_{j} = V_w^\dagger,
\end{equation}
where the notation $j \in w$ indicates to take a product over Pauli $X$ operators acting only on sites where $w$ is in state $1$. Then
\begin{eqnarray}
    V_{w}G_{w}V_{w}^\dagger &=& G_{0}.
\end{eqnarray}
Since $G_{0}$ and $G_{w}$ are unitarily equivalent, their eigenvalues are identical. We will therefore focus on target state $|0\rangle^{\otimes L} \equiv |\textbf{0}\rangle$ for the remainder of the work.

The Grover operator $G_{0}$ has two eigenstates with nonzero quasi-energies, which we will refer to as the \emph{special states},
\begin{eqnarray}
    G_{0}\frac{1}{\sqrt{2}}(|\textbf{0}\rangle\pm i |\Bar{x}\rangle) &=& [\frac{2}{N}-1\mp i\frac{2\sqrt{N-1}}{N}] \frac{1}{\sqrt{2}}(|\textbf{0}\rangle\pm i |\Bar{x}\rangle)\nonumber\\
    &=& e^{-i\phi_{F\pm}}\frac{1}{\sqrt{2}}(|\textbf{0}\rangle\pm i| \Bar{x}\rangle),\label{special state energies}
\end{eqnarray}
where \begin{equation}
    \phi_{F_{\pm}} = \pm(\pi-\arctan\left(\frac{2\sqrt{N-1}}{N-2}\right)) =  \pm(\pi- 2
    \theta),\label{non-zero quasi-energies}
\end{equation}
and
\begin{equation}
    |\Bar{x}\rangle \equiv |\Bar{x}(w=\textbf{0})\rangle = \frac{1}{\sqrt{N-1}} \sum^{N-1}_{j=1} |j\rangle. 
\end{equation}
The rest of the eigenstates, which we will refer to as the \emph{bulk} states, are all degenerate at zero quasi-energy,
\begin{equation}
    G_{0}|\Bar{x}_{n}\rangle = |\Bar{x}_{n}\rangle\label{eq:G0-action-on-bulk-quasienergies}.
\end{equation} 
Noting that $|\Bar x\rangle$ is a ``zero-momentum'' superposition of the nontarget states, we choose the following basis for the degenerate bulk subspace:
\begin{equation}
    |\Bar{x}_{n}\rangle = \frac{1}{\sqrt{N-1}}\sum_{j=1}^{N-1} e^{i(j-1)k_{n}}|j\rangle,
\end{equation}
where $k_{n} = (2\pi n)/(N-1)$, and $n = 1,2,\cdots,N-2$. In the absence of noise, these states are gapped out \emph{i.e.} there is an energy difference (gap) between these bulk states and the states involved in the dynamics, and the evolution of the system takes place in the subspace that does not involve these states.

\subsection{Gate decomposition}

For any practical realization of Grover's algorithm, the Grover operator $G_{0}$ must be decomposed into one- and two-qubit gates. This can be done for any unitary operator with an appropriate universal set of quantum gates \cite{nielsen_chuang_2010}, but the decomposition is not unique. Therefore, choosing an efficient gate decomposition will be crucial to the performance of algorithm upon adding noise.

\begin{figure}
    \centering
\includegraphics[width=0.45\textwidth]{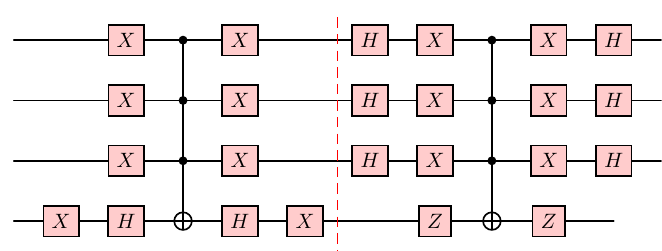}
\caption{Circuit for the four qubit Grover operator $G_{0}$.  $X, Z$ and $H$ are the Pauli $X$, Pauli $Z$ and the Hadamard gate respectively.}
    \label{fig:Grover circuit for L = 5.}
\end{figure}
\begin{figure}
    \centering
\includegraphics[width=0.48\textwidth]{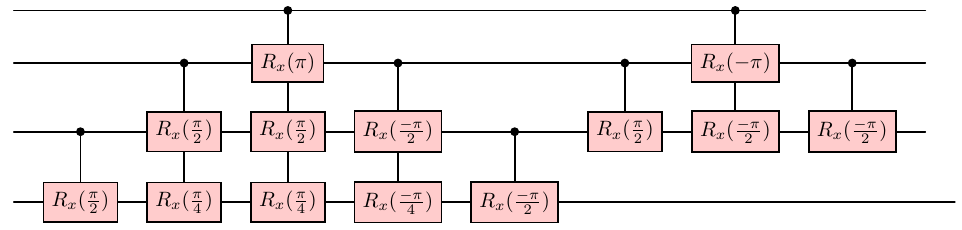}
    \caption{Example decomposition of the four-qubit-controlled $X$ gate in Fig \ref{fig:Grover circuit for L = 5.} in linear depth using the algorithm described in \cite{PhysRevA.87.062318}. The gates with the same control qubit are shown together.}
    \label{fig:Linear depth decomposition for L = 4.}
\end{figure}

In Figure \ref{fig:Grover circuit for L = 5.}, a quantum circuit for Grover's operator $G_{0}$ for four qubits is shown. The oracle and the Grover's diffusion operators are written in terms of a multi-controlled $X$, a.k.a. $L-$qubit Toffoli, gate for ease of decomposition, as such gates are more commonly studied in the literature than the desired multi-controlled $Z$ gate. Decomposing the $L$--qubit Toffoli can be done in many different ways, such as using automated transpilation methods (e.g. `transpile' in Qiskit) or via the general algorithm described in the classic paper by Barenco \emph{et al}.~\cite{PhysRevA.52.3457}. However, these decompositions result in circuits whose depth scales nonoptimally with system size; it appears to scale exponentially with $L$ using Qiskit transpilation and is known to scale as $L^2$  using the Barenco algorithm. In anticipation of improving performance on the noisy algorithm, we instead use the linear depth decomposition algorithm described in \cite{PhysRevA.87.062318}. For a multi-controlled $X$ gate with $L$ qubits, this algorithm provides a decomposition with circuit depth $8L-20$ and with $2L^{2}-6L+5$ controlled $x$-rotation gates\footnote{For a rigorous argument for the linear depth of this decomposition, see \cite{PhysRevA.87.062318, PhysRevA.106.042602}}, hereafter denoted by $CR_{x}(\theta)$.  An example for the case of $L=4$ is shown in Fig \ref{fig:Linear depth decomposition for L = 4.}.  Note that, while the depth is linear, it relies on the ability to perform multiple finite-range \footnote{The decomposition shown requires gate range proportional to $L$, but the authors argue that finite-range gates still allow depth $O(L)$.}  control gates in parallel. Note also that here a single Floquet unitary already has extensive depth $\sim L$, as opposed to more conventional Floquet operators whose depth is usually $L$-independent.\par

\subsection{Systematic noise}

We consider a coherent systematic noise model with random under or over-rotation of each single or two-qubit quantum gate,
\begin{equation}
    U_{k}(\theta_{k}) \to U_{k}(\theta_{k}+\delta \epsilon_{k}),\label{systematic-noise-model}
\end{equation}
where the index $k$ refers to the $k$th gate in the decomposition of the Grover operator, $\epsilon_{k}$ is drawn from the uniform distribution $[-1,1]$ and $\delta$ is the noise strength. Importantly, each Floquet cycle has an identical noise configuration, so that the operator $G_{w}$ is identical across time steps. This ensures that each iteration of the Grover algorithm is a (random) Floquet problem.

To implement this systematic noise model for our one-qubit gates, we start from unitaries of the form $U_k(0) \equiv \sigma_{k} \in \{X, Z, H\}$  which represent single-qubit $\pi$  rotations around the $\hat{x}$, $\hat z$, and $(\hat x + \hat z) / \sqrt 2$ axes, respectively. We can write this as
\begin{equation}
    U_k(0) = R_{\sigma_{k}}(\pi),\mbox{ where } R_{\sigma_{k}}(\theta) = \exp\{-\frac{i\theta}{2}(I-\sigma_{k})\}
\end{equation}
represents a rotation by angle $\theta$ about the $\sigma_{k}$  axis.  Then we define the noisy gates as a random over -or under-rotation, 
\begin{equation}
     U_{k}(\delta \epsilon_{k}) = R_{\sigma_{k}}(\pi+2\delta \epsilon_{k}).\label{eq:adding-noise-in-single-qubit-gates}
\end{equation}
Similarly, we write the noisy two-qubit controlled rotations as
\begin{equation}
    CR_{x}(\theta_{k}+\delta\epsilon_{k}) = |0\rangle\langle 0|_{c}\otimes I_{t} + |1\rangle \langle 1|_{c}\otimes R_{x}(\theta_{k}+\delta\epsilon_{k})_{t}\label{controlled-Rx-gate}
\end{equation}
where the convention is chosen such that $CR_x(\pi)=|0\rangle\langle 0|_{c}\otimes I_{t} + |1\rangle \langle 1|_{c}\otimes X_t$. Note that an identical noise amplitude is added to each control gate independent of the magnitude of the desired rotation $\theta_k$. For small $\theta_k \lesssim \delta$, this means that the gate is predominately noise, suggesting that further improvements can be made by simply neglecting gates that fall below this threshold. Doing so would complicate the analysis below, but will be worth investigating in future research.

\section{Results}

\subsection{Spectrum of noisy Grover operator}

To begin exploring the effect of system noise on the performance of the algorithm, we numerically calculate the Floquet quasi-energies and eigenstates of the Grover operator in the presence of systematic noise. To probe the thermalization of the bulk states, we calculate the bipartite von Neumann entanglement entropy $S_A$ of each eigenstate. We choose a subsystem $A$  of $L/2$  consecutive qubits (with periodic boundary conditions) and calculate the entropy $S_{A} = -\text{Tr}(\rho_{A}\ln \rho_{A})$, where $\rho_{A} = \text{Tr}_{\Bar A}(|\psi\rangle \langle \psi|)$  is the reduced density matrix in eigenstate $|\psi\rangle$. For each eigenstate, the entanglement entropy is averaged over the $L/2$  independent choices of subsystem $A$.

\begin{figure*}
\includegraphics[width=0.9\textwidth]{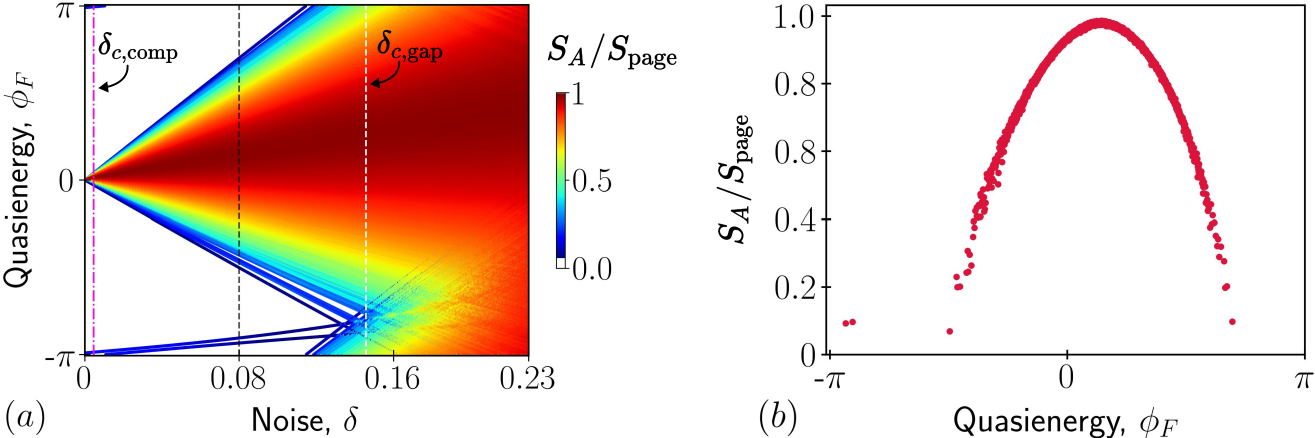}
\caption{(a)  Quasienergies $(\phi_{F})$ of the Grover operator as a function of noise strength $(\delta)$ for $L=12$. The colorbar represents the average bipartite von Neumann entanglement entropy $S_A$ normalized by its Page value \cite{PhysRevLett.71.1291}, $S_{\rm Page}=3.66$ for $L=12$. The data are shown for a single realization of systematic noise $\{\epsilon_k\}$, which is kept constant for different $\delta$ values. The two critical noise values $\delta_{c,\mathrm{comp}}$ and $\delta_{c,\mathrm{gap}}$ are shown with vertical lines.(b) Entanglement entropy of the bulk quasienergies at noise strength $\delta = 0.08$.}
\label{fig:quasienergy-disorder-plot}
\end{figure*}

An example of the quasi-energies and their entanglement entropies are shown for a single noise realization in Figure  \ref{fig:quasienergy-disorder-plot}. The bulk quasi-energies have their degeneracy broken by the noise. Moreover, the quasienergies of these bulk states show a clear linear dependence on the noise strength $\delta$, which comes from the exponential degeneracy of the unperturbed spectrum. This suggests that the bulk properties are well-described by first-order perturbation theory, even at relatively large $\delta$, which we use in the next section to derive an effective Hamiltonian for the quasi-energy spectrum.

In addition to this linear $\delta$-dependence of quasi-energy, the bulk eigenstates show entanglement characteristic of a thermal quantum system, where they should match the thermal entropy $S(E)$. The entanglement peaks in the middle of the bulk spectrum at approximately its maximum value $S_\mathrm{Page}=\left[L\ln(2) - 1\right]/2$ \cite{PhysRevLett.71.1291}. As expected, the entropy density approaches zero at either edge of the bulk spectrum. Further evidence for thermalizing (random matrix) behavior will be given in Sec \ref{sec:rmt_H_eff}.

Finally, below some critical value of noise strength $\delta < \delta_{c,\rm gap}$, the special states remain gapped from the bulk. Since Grover's algorithm involves coherent oscillations within the special state manifold, this suggests that there is a ``computing'' regime at small $\delta$  where finite probability of finding the target state will survive the systematic noise.  We will indeed show that this is true at small enough $\delta$ though, as we will see in our analysis of the special states (Sec \ref{sec:special_states}), the computing transition happens at a much smaller critical value $\delta_{c,\rm comp}\ll\delta_{c,\rm gap}$. Anticipating the results that we are about to show, $\delta_{c,\rm gap}$ will in fact correspond to an ergodicity transition; 
for $\delta < \delta_{c,\mathrm{gap}}$, the initial states within the special states subspace do not reach their thermal equilibrium expectation values \emph{i.e.} retain memory of their initial conditions even after a long time, while for $\delta > \delta_{c,\mathrm{gap}},$ all states obtain their thermal expectation values.

\subsection{Effective Hamiltonian of the bulk}

\begin{figure}
\centering
\includegraphics[width=0.49\textwidth]{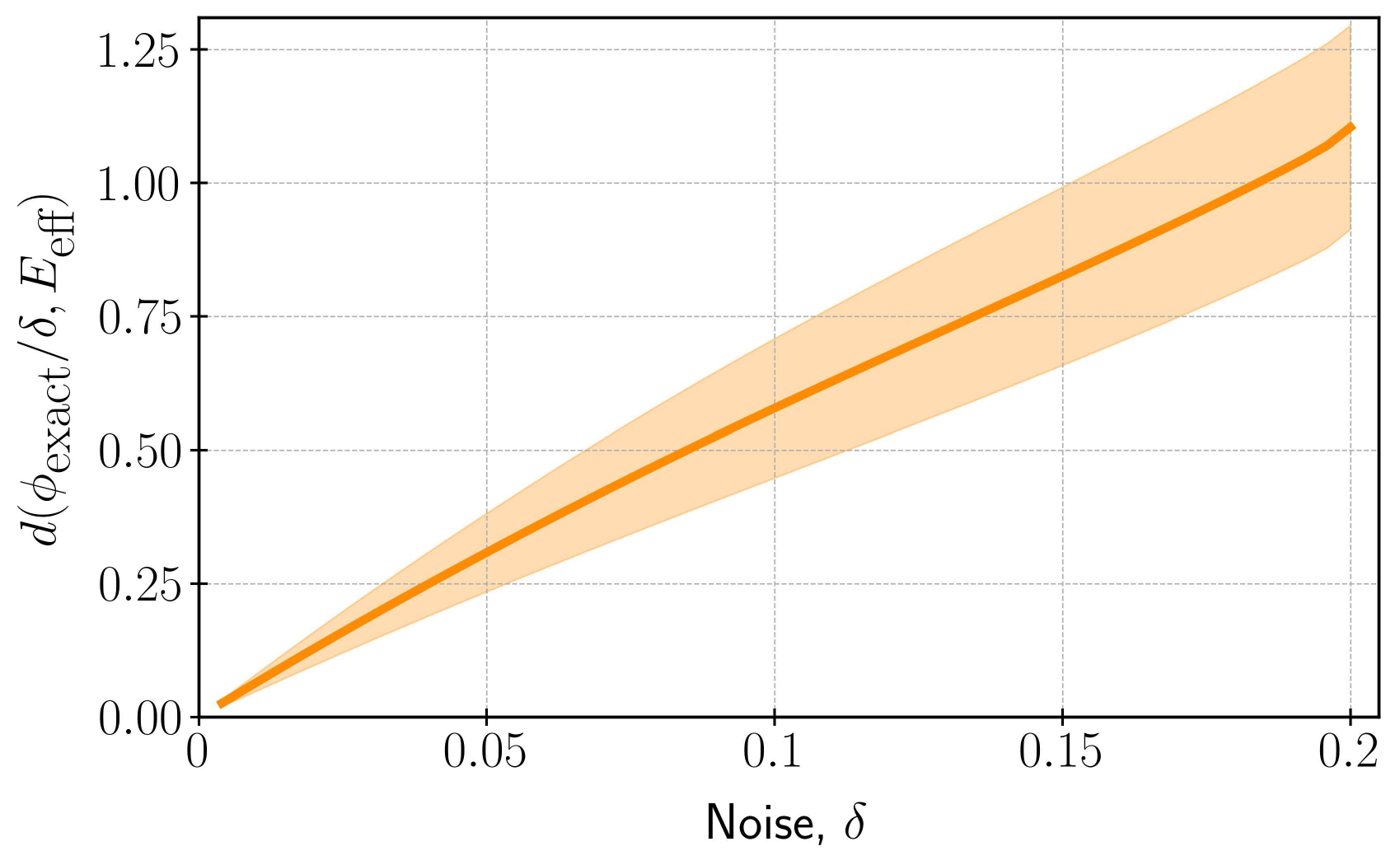}
\caption{Plot of noise averaged difference between the exact quasi-energy  of $G(\delta)$ and eigenvalues of $\delta H_{\rm eff}$ for $L=8$ using Eq. \eqref{eq:exact-effective-difference-bulk}. Error bands represent the standard error of mean for five noise realizations.}
\label{fig:exact vs first order}
\end{figure}

We start by using our observation of clear linear $\delta$-dependence of the bulk spectrum to derive an effective Hamiltonian from first-order degenerate perturbation theory. To achieve this goal, let us begin by writing the Grover operator with noise strength $\delta$ as
\begin{equation}
    G(\delta) = e^{-iH_{M}(\theta_{M}+\delta \epsilon_{M})}\cdots e^{-iH_{1}(\theta_{1}+\delta \epsilon_{1})},
\end{equation}
where we have written each gate in the decomposition of the Grover operator as generalized rotations in Hilbert space [see Eq. \eqref{eq:adding-noise-in-single-qubit-gates}]. Here $M = 4L^{2}-6L+10$ is the total number of gates in the decomposition of the $L$-qubit Grover operator. Individual commuting gates are divided into separate terms in the product. Expanding the exponentials and keeping terms only up to first order in $\delta$, we obtain:
\begin{eqnarray}
    G(\delta)  &\approx& e^{-iH_{M}\theta_{M}}(1-i\delta\epsilon_{M}H_{M})\cdots e^{-iH_{1}\theta_{1}}(1-i\delta\epsilon_{1}H_{1})\nonumber\\
    &\approx& (1-i\delta H_{\rm eff})G_{0},\label{eq:H_eff-within-first-order }
\end{eqnarray}
where we have defined the effective Hamiltonian as
\begin{widetext}
\begin{equation}
    H_{\rm eff} = \sum_{k=1}^{M} \epsilon_{k}e^{-i H_{M}\theta_{M}}e^{-i H_{M-1}\theta_{M-1}}\cdots e^{-i H_{k+1}\theta_{k+1}}H_{k}e^{i H_{k+1}\theta_{k+1}}\cdots e^{i H_{M-1}\theta_{M-1}}e^{i H_{M}\theta_{M}}. \label{eq:effective-Hamiltonian-definiton}
\end{equation}
\end{widetext}
We identify $H_{\rm eff}$ as the generator of gate perturbations in the Heisenberg picture \footnote{Here, the origin of time in the Heisenberg picture is defined as the end of the Floquet cycle, such that Heisenberg evolution ``rewinds'' the action of $e^{-iH_M \theta_M} e^{-iH_{M-1} \theta_{M-1}} \cdots$}. To demonstrate that $H_{\rm eff}$ correctly describes the exact bulk quasienergies, let us consider the action of $G(\delta)$ on these eigenstates. Using Eq. \eqref{eq:G0-action-on-bulk-quasienergies}, and keeping terms within the first order in $\delta$, we get
\begin{eqnarray}
    G(\delta)|\Bar{x}_{n\neq 0}\rangle &\approx& (1-i\delta H_{\rm eff})G_{0}|\Bar{x}_{n\neq 0}\rangle,\nonumber\\
    &=& (1-i\delta H_{\rm eff})|\Bar{x}_{n\neq 0}\rangle,\nonumber\\
    &\approx& e^{-i\delta H_{\rm eff}}|\Bar{x}_{n\neq 0}\rangle\label{eq:H_eff-within-first-order}
\end{eqnarray}
Thus, after discarding the two special states from $ H_{\rm eff}$, its eigenvalues should match with the exact bulk quasienergies of $G(\delta)$. In Fig \ref{fig:exact vs first order}, we plot the noise-averaged difference between the exact and the effective energies:
\begin{equation}
    d\left(\frac{\phi_{\rm exact}}{\delta},E_{\rm eff}\right) \equiv \left[\sum_{k \in \rm bulk}\left(\frac{\phi_{\rm exact, k}}{\delta}-E_{\rm eff, k}\right)^{2} \right]^{1/2} \label{eq:exact-effective-difference-bulk}
\end{equation} 
as a function of noise strength $\delta$. For small $\delta$, the $k-$th quasienergy can be written as $\phi_{\textrm{exact},k} = \delta E_{\textrm{eff},k} + \mathcal{O}(\delta^{2})$. From this, we can write 
$\phi_{\textrm{exact},k}/\delta - E_{\textrm{eff},k} = \mathcal{O}(\delta^{2})/\delta$. Then plotting $\phi_{\textrm{exact},k}/\delta - E_{\textrm{eff},k}$ as a function of $\delta$ will asymptote to zero as $\delta \to 0$ if the higher order terms $\delta^{3}, \delta^{4}$ are small. In Fig \ref{fig:exact vs first order} we plot the disorder averaged sum of all these $k-$quasienergies. We see that deviations vanish linearly as $\delta \to 0$, confirming the correctness of first-order perturbation theory. Indeed, we numerically see that first-order perturbation accurately captures the spectrum nearly up to the value of $\delta$ when the bulk spans the entire Floquet Brillouin zone from $-\pi$ to $\pi$.
It is worth noting that our effective Hamiltonian is defined to be independent of the scaling parameter $\delta$, which allows us to analyze the statistical properties of $H_{\rm eff}$ for different noise strengths and therefore determine many important properties such as thermalization of the bulk states and scaling of $\delta_{c,\rm gap}$. Having shown the accuracy of  $H_{\rm eff}$  in modeling our spectrum, we now turn to analyzing its properties with a focus on the bulk states. $H_{\rm eff}$  consists of random terms which start as 2-local terms before Heisenberg evolution causes operator spreading. Some remnant of the locality remains, as we argue numerically in Appendix \ref{sec:matrix_elements_of_Heff}. However, lacking any other structure, we would expect this Hamiltonian to exhibit generic, thermalizing behavior. 
An exact analytical treatment of this Hamiltonian is not feasible, therefore we resort to analyzing its properties statistically using random matrix theory, which is a natural and powerful framework to study disordered many-body systems. One way to test if the Hamiltonian belongs to a random matrix ensemble is to calculate its energy level statistics and the Kullback-Leibler divergence that captures the level repulsion which is a characteristic feature of a thermalized phase. On the other hand, if the energy level statistics is close to the Poisson distribution, the Hamiltonian will display properties of a nonergodic system. We now show that $H_{\textrm{eff}}$ thermalizes by numerically demonstrating its consistency with random matrix theory.
\subsubsection{Random matrix theory treatment of $H_{\mathrm{eff}}$ \label{sec:rmt_H_eff}}
In the thermal phase, the energy levels of a many-body Hamiltonian show level repulsion, and their statistics follow that of a random matrix, the underlying ensemble of which depends on the symmetries of the Hamiltonian in the problem. Here, we examine two properties of $H_\mathrm{eff}$  to demonstrate signatures of random matrix theory (RMT). First, to indicate the existence and amount of level repulsion (which is characteristic of the RM ensemble), we examine the level spacing ratio $r_{n} = \text{min}(\Delta E_{n},\Delta E_{n+1})/\text{max} (\Delta E_{n},\Delta E_{n+1})$, where $\Delta E_{n} = E_{n}-E_{n-1}$ is difference between the ordered eigenenergies of $H_{\rm eff}$ \cite{PhysRevB.75.155111}. Second, to study the ergodic structure of the eigenstates, we calculate the Kullback-Leibler (KL) divergence (also called the relative entropy) in the computational basis. It is defined as
$\text{KLd}_{n} = \sum_{i=1}^{dim \mathcal{H}}  p_{i} \ln (p_{i}/q_{i})$, where $p_{i} = | \langle i|n\rangle|^{2}$ and $q_{i} = | \langle i|n+1\rangle|^{2}$ are the occupation probabilities of two energetically-neighboring eigenstates $|n\rangle$ and $|n+1\rangle$ in the $Z$ (computational) basis $\{|i\rangle\}$ \cite{10.1214/aoms/1177729694,PhysRevB.91.081103}. To calculate these two metrics, we sample $H_{\rm eff}$ for different noise realizations, and calculate the level spacing ratio and KL divergence for each, and then average over noise realizations.
\begin{figure}
\centering
\includegraphics[width=0.48\textwidth]{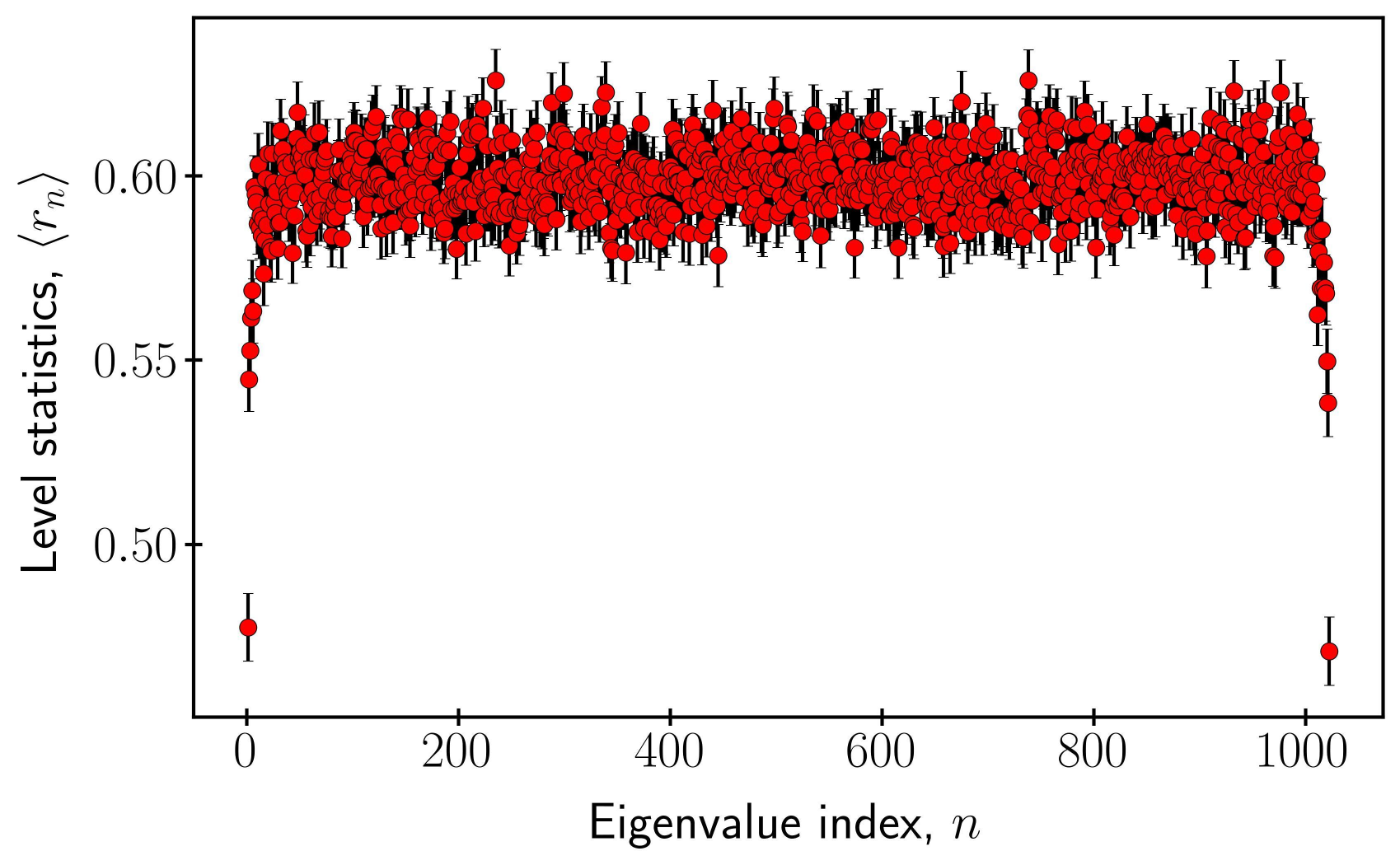}
\includegraphics[width=0.48\textwidth]{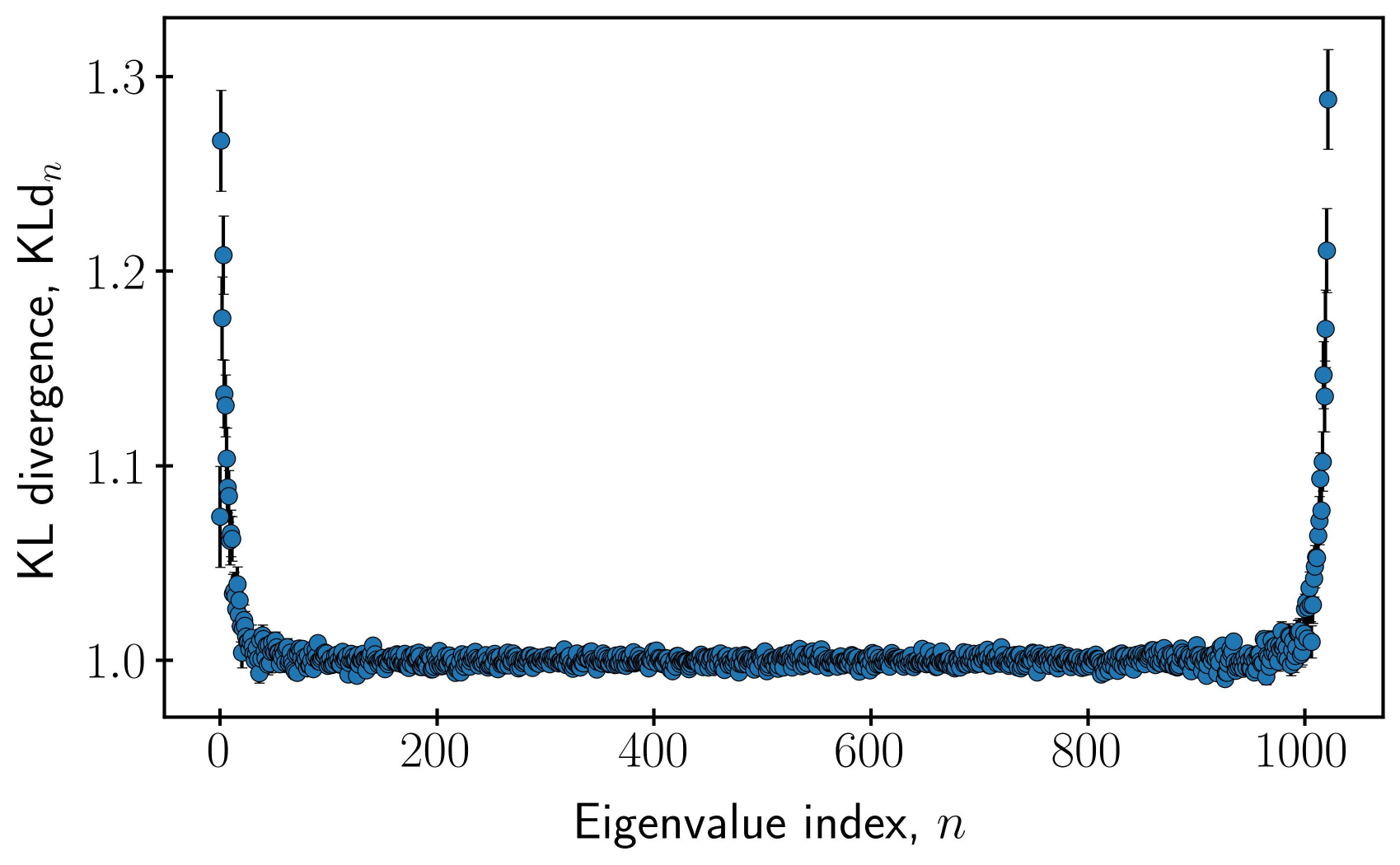}
\caption{Plot of noise-averaged level spacing ratio (top) and Kullback-Leibler divergence (bottom) of $H_{\rm eff}$ for $L=10$. Error bars represent standard error of mean for $768$ noise realizations.}
\label{fig:level-statistics}
\end{figure}
The results, shown in Fig \ref{fig:level-statistics} indicate that the noise-averaged level spacing ratio $\langle r_{n}\rangle$ and KL divergence $\langle \text{KLd}_{n}\rangle$ are consistent with that of Gaussian Unitary Ensemble (GUE): $r_{\rm GUE} = 0.599$\cite{PhysRevLett.110.084101}, $\text{KLd}_{\rm GUE} = 1$\footnote{We calculated the KL divergence of the standard Gaussian Unitary Ensemble (GUE) by generating GUE matrices of size $2^{14}\times 2^{14}$ using the package TeNPy}. This confirms that the system will thermalize within the bulk state manifold that is well-described by $H_\mathrm{eff}$; note, however, that the data shown actually include all states, not just the bulk. Furthermore, GUE indicates that thermalization will occur with no conserved quantities besides energy, as is perhaps expected from the unstructured form of $H_\mathrm{eff}$. Note, however, that both the level spacing ratio and K-L divergence probe only properties of states that are nearby one another in energy. Therefore, other RMT properties that rely on the complete spectrum and full random matrix form -- such as the Wigner semicircle law -- are not in general expected to hold. In fact, we will see that the locality structure that exists in $H_\mathrm{eff}$  prevents the formation of a Wigner semicircle, as is known to happen in other local ergodic Hamiltonians \cite{rodriguez2023quantifying}.

\subsection{Critical noise strength and its scaling \label{sec:critical_disorder}}

Having shown that $H_\mathrm{eff}$ obeys certain properties of RMT, we now use that to solve for other properties of the noisy algorithm, starting with the critical noise $\delta_{c,\rm gap}$  and its scaling with system size. We start by noting that the algorithm stops giving meaningful answers once the two special states enter into the bulk, i.e., once the gap closes. As evident from Fig \ref{fig:quasienergy-disorder-plot}, this happens near the point where the bulk energies span from $-\pi$ to $\pi$ due to weak dependence of the special state energies on $\delta$. We will study $\delta$-dependence of the special state energies in Sec \ref{sec:special_states}. We define the noise value at this gap-closing point as the critical noise $\delta_{c,\rm gap}$. As indicated by our numerical findings, the quasi-energy of the noisy Grover operator is well captured by the effective Hamiltonian up to the point of merger of the special states, therefore we will study the eigenvalues of the latter to determine the critical noise and its scaling with system size.

\begin{figure}
    \centering
    \includegraphics[width=0.45\textwidth]{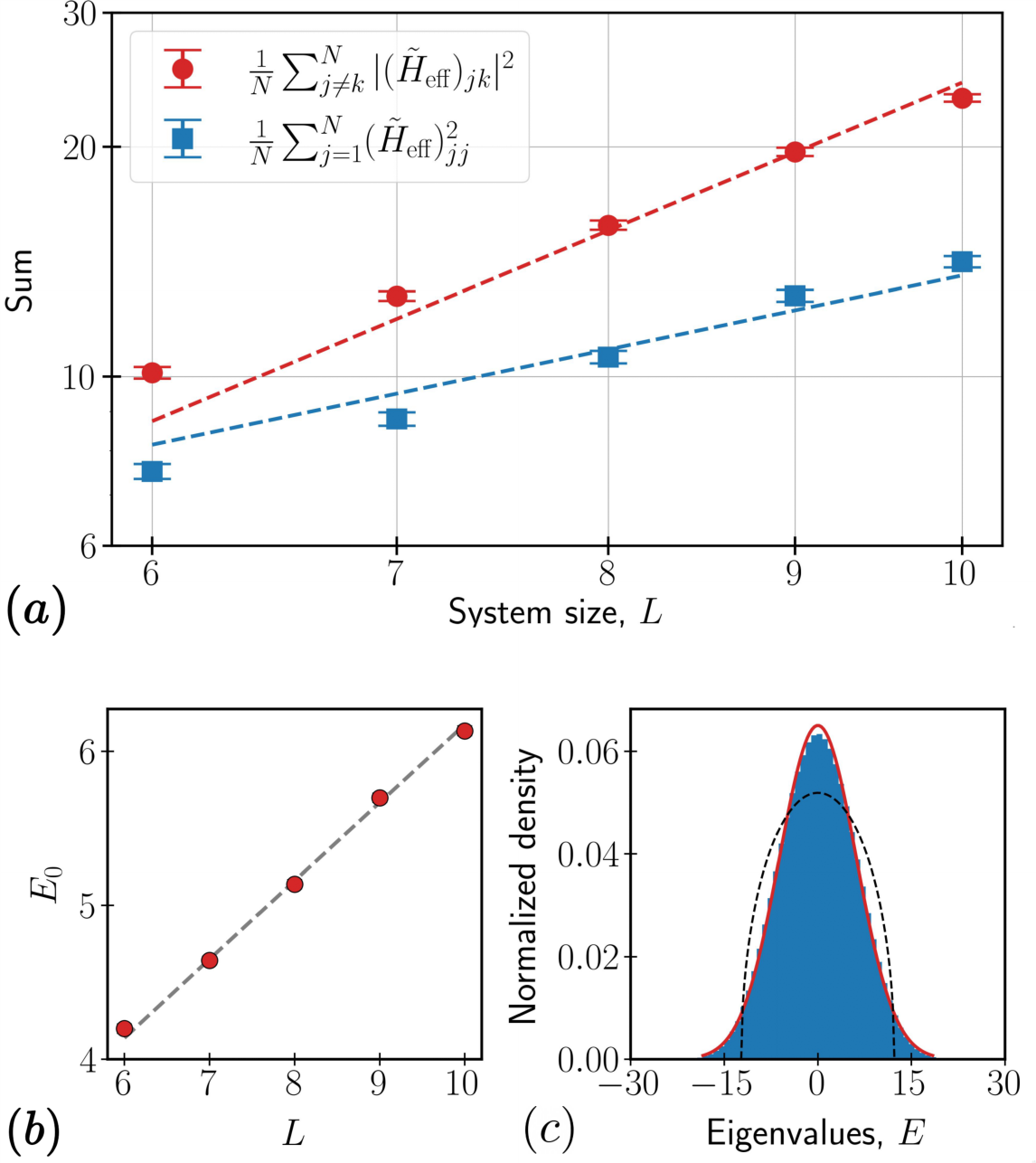}
    \caption{(a) Plot of the sum of squared diagonal and off-diagonal elements of $\Tilde{H}_{\rm eff}$ as a function of system size in a log-log plot. Error bars indicate standard error of mean over $512$ noise realizations. The best-fit functions $f(x) = aL^{2}$  (off-diagonal) and $g(L) = bL$ (diagonal) are shown with red and blue dashed lines respectively. (b) $E_{0}$ as a function of system size $L$ calculated using Eq. \eqref{eq:definition-of-E0} (red) by averaging over $512$ noise realizations, and is compared with the theoretical prediction given by \eqref{eq:E_0_exact}. (c) Distribution of bulk eigenvalues of $H_{\rm eff}$ for $L= 10$, shown along with the normal distribution [Eq. \eqref{eq:normal-density-of-eigenvalues}] (red) and Wigner's semicircle (dashed).}
    \label{fig:sum-of-squared-diagonal-and-off-diagonal-elements_E0_scaling}
\end{figure}

The relevant property of $H_\mathrm{eff}$ that determines $\delta_{c,\rm gap}$ is the second moment of its energy distribution. While each realization of $H_\mathrm{eff}$ will have nonzero trace (first moment), the fact that $H_\mathrm{eff}$ is generated by Paulis implies that this term must be proportional to the identity matrix and therefore creates a uniform energy shift that does not affect the gap-closing transition \footnote{The trace is explicitly given by $\frac{\text{tr}(H_{\rm eff})}{N} = \sum_{k\in\{CR_{x}\}}\frac{\epsilon_{k}}{4}+\sum_{l\in\{Z,H,X\}}\epsilon_{l}$, whose noise-averaged value is equal to zero, while its standard deviation is extensive: $\text{std}\left[\frac{\text{tr}(H_{\rm eff})}{N} \right] \sim L$ given $O(L^2)$ gates}. To determine the second moment, it is convenient to first shift the spectrum to make $H_\mathrm{eff}$ traceless:
\begin{equation}
    \tilde H_{\rm eff} \equiv H_{\rm eff} - \frac{1}{N}\text{Tr}(H_{\rm eff})\mathbf{1},
\end{equation}
which can be done by individually removing the traces from the generators, $\tilde H_j = H_j - \mathrm{Tr} \left( H_j \right) / N$. Then the variance is given by
\begin{align}
E_{0}^{2} & \equiv\mathcal{E}\left\{ \frac{1}{N}\text{Tr}\left[\Tilde{H}_{\text{eff}}^{2}\right]\right\} \label{eq:definition-of-E0}\\
 & =\mathcal{E}\left\{ \frac{1}{N}\sum_{j,k}\epsilon_{j}\epsilon_{k}\text{Tr}\left[\Tilde{H}_{j,\Tilde{H}}\Tilde{H}_{k,\tilde{H}}\right]\right\} \\
 & =\frac{1}{3N}\sum_{j}\text{Tr}\left[\tilde{H}_{j}^{2}\right]\\
 & = \frac{1}{8}(2L^{2}+10L+5), \label{eq:E_0_exact}
\end{align}
where we used that $\mathcal{E}\left(\epsilon_{j}\epsilon_{k}\right)  =\frac{1}{3}\delta_{jk}$. Notice that $E_0 \sim L$ comes from the overall gate count, $M \sim L^2$; by reducing scaling of the gate count, we would modify this system-size dependence. This scaling is corroborated numerically in Fig \ref{fig:sum-of-squared-diagonal-and-off-diagonal-elements_E0_scaling}.
\par
\begin{figure}
    \centering
\includegraphics[width=0.48\textwidth]{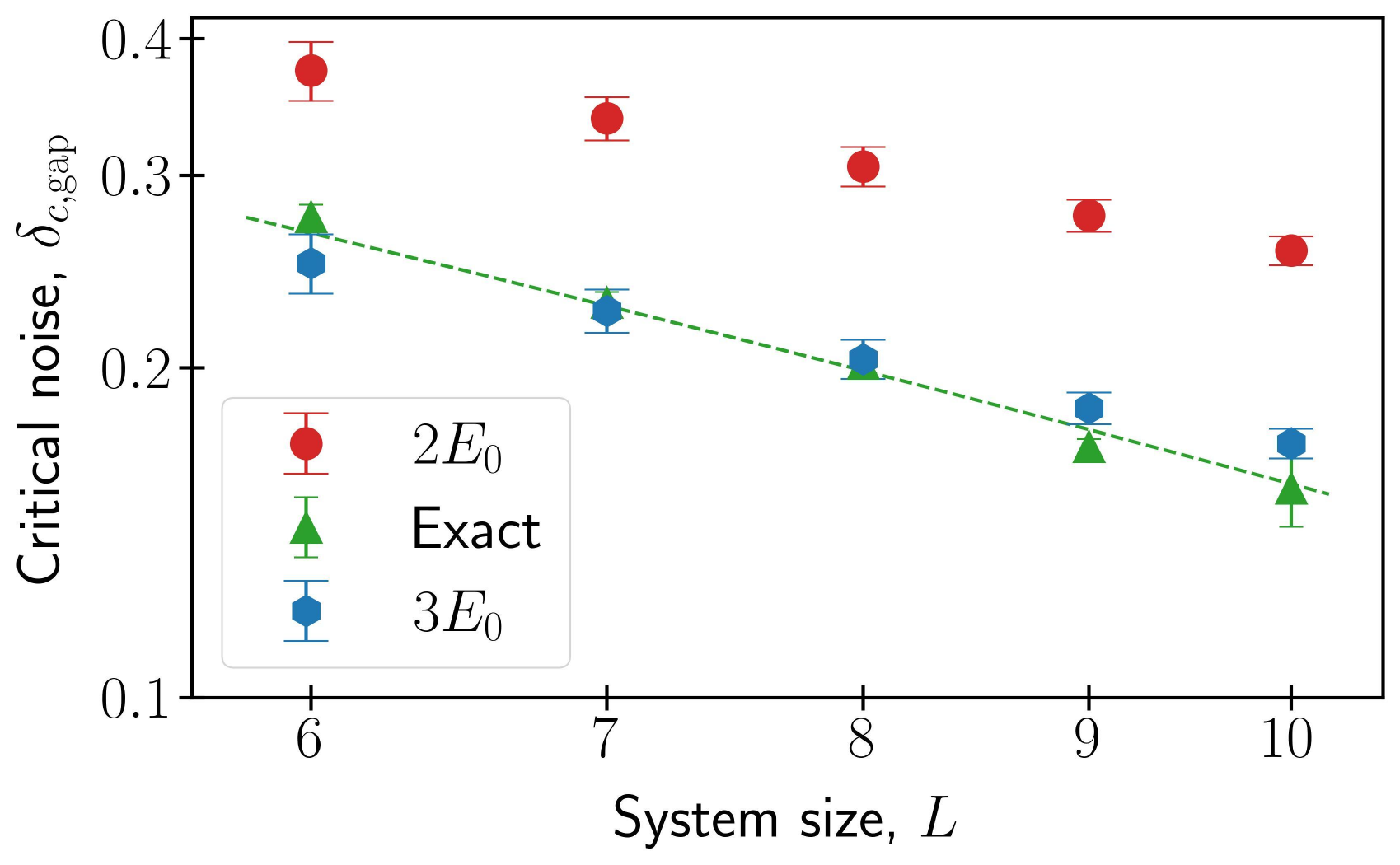}
    \caption{Critical noise strength $\delta_{c,\rm gap}$ as a function of system size shown in a log-log plot. The exact critical noise is calculated at the point where the gap between the two special states and the bulk closes and is averaged over $20$ noise realizations. The dashed line on the plot represents the best-fit line for these exact noises. Predicted critical noises are averaged over $512$ noise realizations and are calculated when the two special states merge into  $67\%, (-2E_{0},2E_{0})$, and $99.7\%, (-3E_{0},3E_{0})$ of the bulk eigenvalues spanning the Floquet Brillouin zone.}
    \label{fig:critical-disorder-exact-effective}
\end{figure}

Having obtained an accurate description of the bulk spectrum, we can derive the scaling of the critical noise $\delta_{c,\rm gap}$. Since the Gaussian distribution does not have a sharp cutoff, let us instead consider when the eigenvalues within, say, $3\sigma$ span the entire Floquet Brillouin zone. Then we can write
\begin{eqnarray}(-3\delta_{c,\rm gap}E_{0},3\delta_{c,\rm gap}E_{0}) = (-\pi,\pi),
\end{eqnarray}
which gives
\begin{eqnarray}
    3\delta_{c,\rm gap}E_{0} = \pi \implies \delta_{c,\rm gap} = \frac{\pi}{3E_{0}} \implies \delta_{c,\rm gap} \sim L^{-1}.\nonumber\label{eq:critical-disorder-and-its-scaling}
\end{eqnarray}
It is easy to see that irrespective of the fraction of eigenvalues we consider that span the entire Brillouin zone, the scaling of the critical noise does not change. In Fig.\ref{fig:critical-disorder-exact-effective} we illustrate this scaling by considering $2\sigma$ and $3 \sigma$ of the bulk eigenvalues. The parameter $\delta_{c, \mathrm{gap}}$ does not represent a conventional phase transition since it scales as a power law with system size $L$. However, one can define the observable: the late-time-average of occupation in the state $|\Bar{x}\rangle$, which vanishes at $\delta = \delta_{c,\mathrm{gap}}$ in a manner characteristic of conventional second-order phase transitions. An alternative perspective is to consider $\delta_{\mathrm{eff}} = L \delta$, which represents the aggregate error rate across all qubits. In terms of this rescaled noise parameter, the transition occurs at $\delta_{\textrm{eff}} = \mathcal{O}(1)$. This type of parameter rescaling is common in both mean-field theory models and nonstandard thermodynamic limits \cite{PhysRevB.99.134305}.
\par
Our analysis gives the correct system-size scaling of $\delta_{c,\rm gap}$, but also shows that a full random matrix treatment is incorrect.  Specifically, the drastically different variances of the diagonal and the off-diagonal elements of $H_{\rm eff}$ in Fig.\ref{fig:sum-of-squared-diagonal-and-off-diagonal-elements_E0_scaling} deviate from the prediction of a Gaussian Unitary Ensemble (GUE):
\begin{equation}
    \mathcal{E}_\mathrm{GUE} \left[ H^{2}_{nn} \right] = \mathcal{E}_\mathrm{GUE} \left[ |H_{mn}|^{2}\right] = \frac{\lambda^{2}}{N},
\end{equation}
for $1\leq m,n\leq N$, and where $\mathcal{E}_\mathrm{GUE} \left[ \cdot \right]$ denotes the average over different realizations of the GUE and $\lambda$ sets the units of energy \cite{Haake_2010}. As a result of this discrepancy in the scaling of elements, the distribution of the eigenvalues of $H_{\rm eff}$ deviates from Wigner's semicircle law, instead following a normal distribution
\begin{equation}
    \rho(E_{\rm eff}) = \frac{1}{E_{0}\sqrt{2\pi}} e^{-(E_{\rm eff}/E_{0})^{2}/2}.
    \label{eq:normal-density-of-eigenvalues}
\end{equation}
Similar deviations from the semicircle law in the energy spectra of many-body systems have been noted in other interacting many-body systems\cite{BOHIGAS1971261,FRENCH1970449,e18100359}, leading to the conclusion that completely random matrices do not accurately represent realistic many body systems. The underlying assumption of these matrices -- that each particle interacts simultaneously with all other particles -- contradicts the nature of most systems where interactions are predominantly few-body. The presence of $k$-locality here is not immediately obvious due to operator spreading of the $2$-local perturbations when taken in the Heisenberg picture (see Eq.~\ref{eq:effective-Hamiltonian-definiton}). We find strong evidence for approximate $k$-locality from a detailed analysis of matrix elements in $H_\mathrm{eff}$, which is done in Appendix \ref{sec:matrix_elements_of_Heff}.


It is worth noting that the exact value of the gap closing between bulk and special states will not happen precisely at $\delta_{c,\rm gap}$ derived above, but rather a slightly smaller value. This is because the edges of the spectrum of $H_\mathrm{eff}$ do not scale identically as its variance. For example, conventional local models have an extensive energy range but have energy width scaling as $\sqrt L$. However,  the energy width $E_0 \sim L$ is the most important parameter for determining ergodicity, as this is the energy scale on which a finite density of states will come in contact with the special states. The precise scaling of the range of bulk energies, which determines initial gap closing but not ergodicity breaking, will be left for future work.

\subsection{Analysis of the special states \label{sec:special_states}}

\begin{figure}
\includegraphics[width=0.48\textwidth]{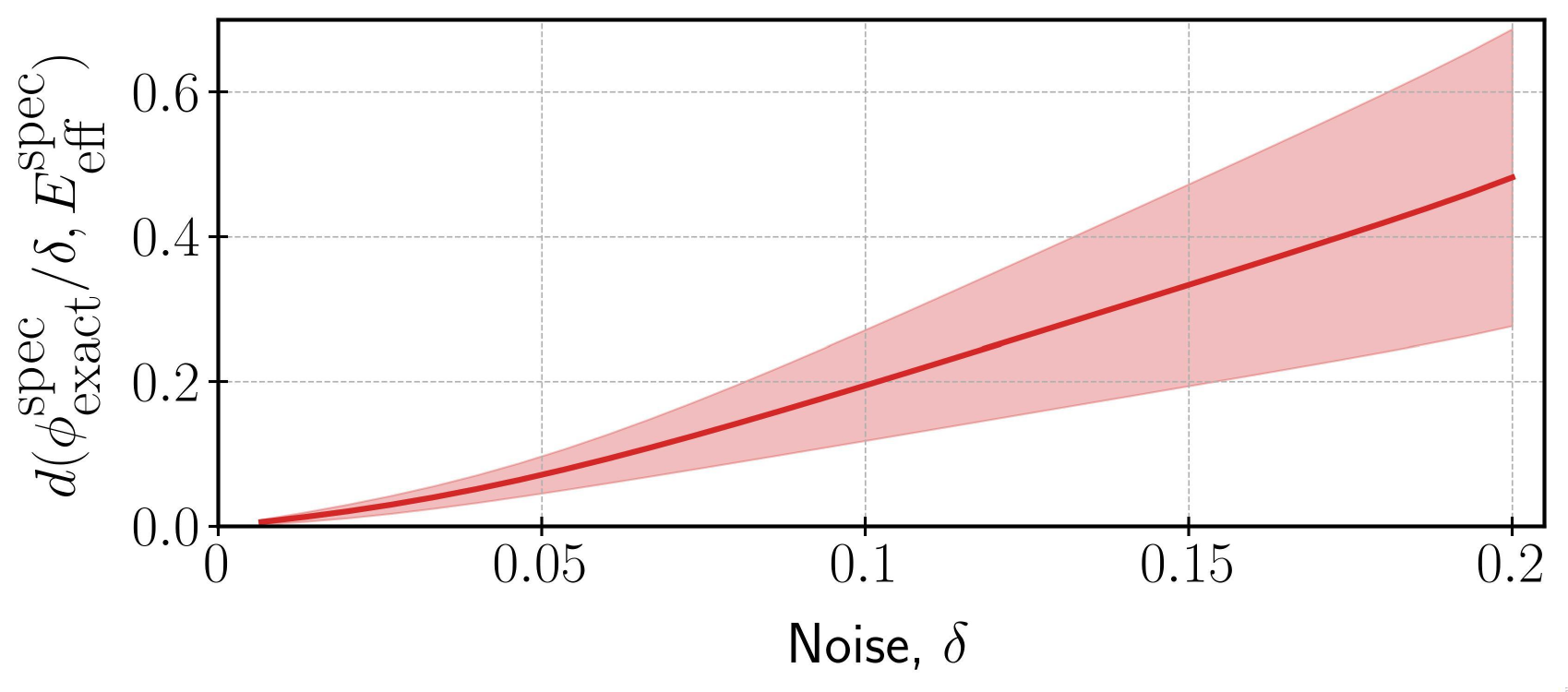}
    \caption{
    Plot of difference between the exact special states quasi-energy  of $G^{\rm spec}(\delta)$ and eigenvalues of $-\theta\tau_{y}+\delta H^{\rm spec}_{\rm eff}$ for $L=8$. Error bands represent the standard error of mean for five noise realizations.}
    \label{fig:exact-first-order-special-states-quasienergies}
\end{figure}

Finally, we turn our attention to the two special states. In the absence of noise, the actions of the Grover operator on the states $|\textbf{0}\rangle$ and $|\Bar{x}\rangle$ are
\begin{eqnarray}
    G_{0}|\textbf{0}\rangle &=& \left(\frac{2}{N}-1 \right)|\textbf{0}\rangle+\frac{2\sqrt{N-1}}{N}|\Bar{x}\rangle\\ &=& -\cos(2\theta) |\textbf{0}\rangle + \sin(2 \theta) |\Bar x \rangle,\\
    G_{0}|\Bar{x}\rangle &=& -\frac{2\sqrt{N-1}}{N}|\textbf{0}\rangle+\left(\frac{2}{N}-1 \right)|\Bar{x}\rangle\\ &=& -\cos(2\theta) |\Bar x \rangle - \sin(2 \theta) |\textbf{0}\rangle.
\end{eqnarray}
Then we can write the noiseless Grover operator in the $\{|\textbf{0}\rangle, |\Bar{x}\rangle\}$ block as
\begin{equation}
    G^{\rm spec}_{0} = -\begin{bmatrix}
\cos(2\theta) & \sin(2\theta)\\
-\sin(2\theta) & \cos(2\theta)
\end{bmatrix} = -e^{2i\theta\tau_{y}}.
\end{equation}
Now, adding back in the systematic noise, we can repeat our analysis using the effective Hamiltonian within first-order perturbation theory and apply it to these special states. Restricting Eq.\eqref{eq:H_eff-within-first-order} to the special states manifold by writing it in $|\textbf{0}\rangle$ and $|\Bar{x}\rangle$ basis, 
\begin{equation}
    H^{\rm spec}_{\rm eff} \equiv \begin{bmatrix}
\langle \textbf{0}|H_{\rm eff}|\textbf{0}\rangle & \langle \textbf{0}|H_{\rm eff}|\Bar{x}\rangle\\
\langle \Bar{x}|H_{\rm eff}|\textbf{0}\rangle & \langle \Bar{x}|H_{\rm eff}|\Bar{x}\rangle
\end{bmatrix},\label{eq:H_eff-spec-Hamiltonian}
\end{equation}
we obtain:
\begin{eqnarray}
    G^{\rm spec}(\delta) &\approx& -(1-i\delta H^{\rm spec}_{\rm eff})e^{2i\theta \tau_{y}}\nonumber \\
    &\approx& \exp\{-i(-2\theta\tau_{y}+\delta H^{\rm spec}_{\rm eff}-\pi )\}.\label{eq:special-states-Hamiltonian-first-order}
\end{eqnarray}
The eigenvalues of $-2\theta\tau_{y}+\delta H^{\rm spec}_{\rm eff}-\pi$ capture the quasi-energies of two special states, as shown in Fig \ref{fig:exact-first-order-special-states-quasienergies}, once again confirming the effectiveness of first-order perturbation theory.

Next, let us analyze the statistical properties of $H^{\rm spec}_{\rm eff}$ to understand the behavior of the two special states in the presence of noise. Taking the logarithm on both sides of Eq. \eqref{eq:special-states-Hamiltonian-first-order}:
\begin{equation}
    -i\ln G^{\rm spec} \approx \pi\mathbf{1}+2\theta \tau_{y}-\delta H^{\rm spec}_{\rm eff},\label{eq:log-of-G_spec}
\end{equation}
we can expand in the Pauli basis to obtain
\begin{multline*}
    \pi\mathbf{1} + 2\theta \tau_{y} - \delta H^{\rm spec}_{\rm eff} = -\delta b_{x}\tau_{x}+(2\theta-\delta b_{y})\tau_{y}-\delta b_{z}\tau_{z}\\
    + (\pi - \delta b_0) \tau_0  
\end{multline*}
where $b_{j} = \text{tr}(H^{\rm spec}_{\rm eff}\tau_{j})/2$ with $\tau_{x,y,z,0}$ representing the Pauli matrices ($\tau_0=\mathbf{1}$). Then the square of the energy gap between the two special states is given by
\begin{equation}
    \Delta^2 \approx 4\left[\delta^{2}b^{2}_{x}+(2\theta-\delta b_{y})^{2}+\delta^{2} b^{2}_{z}\right],\label{eq:special-state-energy-gap-Pauli-coefficients}
\end{equation}
which sets the oscillation frequency between the special states. Averaging over noise configurations and noting that $\mathcal{E} \left[ b_{x,y,z,0} \right] = 0$, we see that the gap is set by their variance:
\begin{equation}
    \mathcal{E} \left[ \Delta^2 \right] \approx 4\left[4\theta^2 + \delta^{2}\left(\mathcal{E} \left[ b^{2}_{x}\right] + \mathcal{E} \left[ b^{2}_{y} \right] + \mathcal{E} \left[ b^{2}_{z} \right] \right)\right].\label{eq:special-states-energy-gap-squared-variance}
\end{equation} 

\begin{figure}
    \centering
\includegraphics[width=0.48\textwidth]{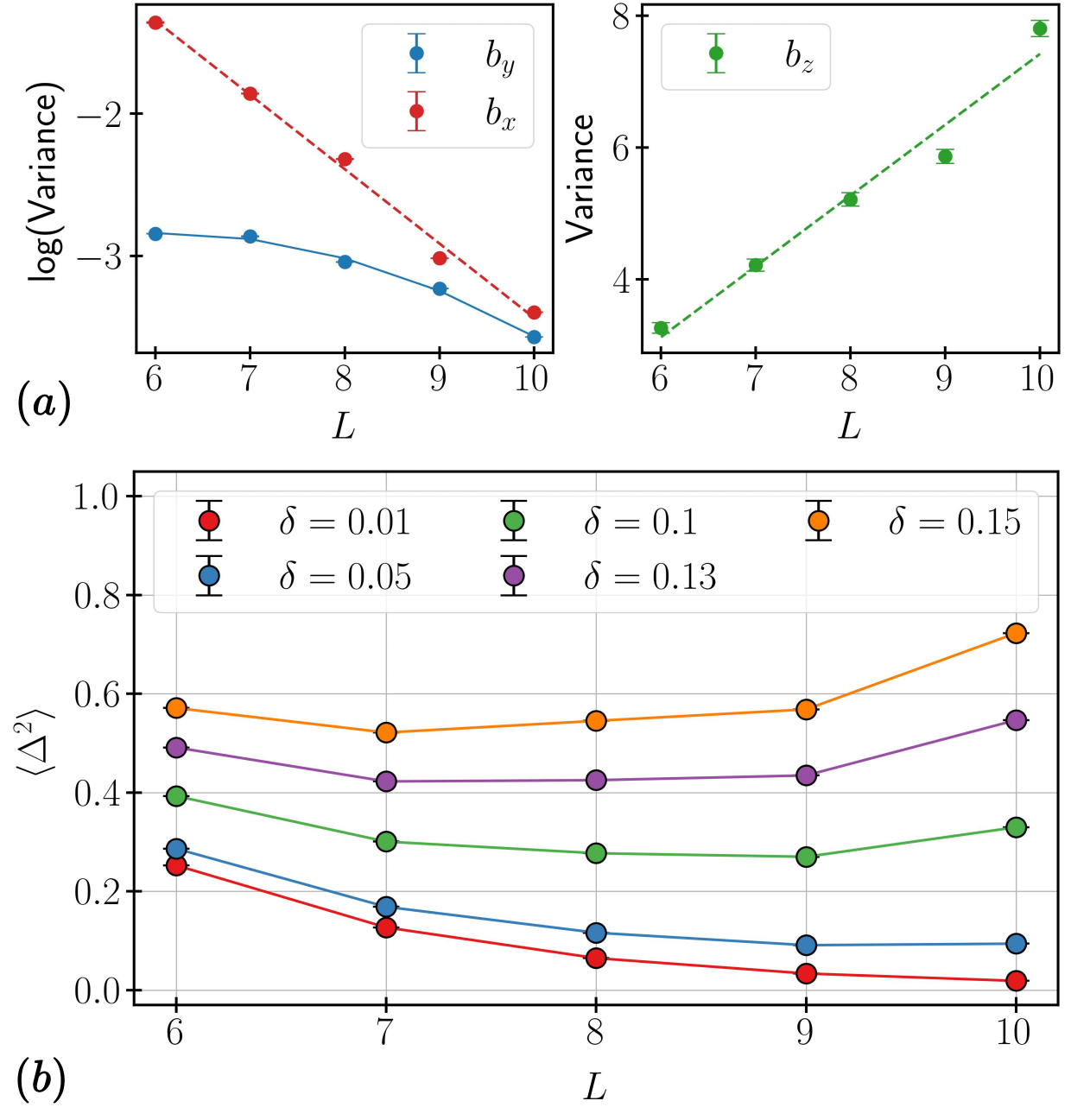}
    \caption{(a) Variance of the coefficients $b_{x},b_{y}$ and $b_{z}$ as a function of system size shown along with their best-fit curves. Data are averaged over $512$ noises. (b) Plot of average energy gap squared $\langle \Delta^{2} \rangle$ obtained using Eq. \eqref{eq:special-state-energy-gap-Pauli-coefficients} as a function of system size $L$ for different noise strengths. Data are averaged over $512$ noise realizations.}
    \label{fig:pauli-coefficients-variance-with-system-size}
\end{figure}
If $H_\mathrm{eff}$ were an $N\times N$ GUE random matrix, then $H^{\rm spec}_{\rm eff}$ would be a $2\times 2$ GUE random matrix, for which $\mathcal E(b_{x,y,z}^2) \sim 1/N$.  However, as we have seen that non-GUE scaling of the diagonals exists in $H_\mathrm{eff}$, we need to more carefully examine whether GUE scaling holds within the special state manifold. As shown in Fig \ref{fig:pauli-coefficients-variance-with-system-size}, indeed we do not have exponential scaling of these Pauli coefficients nor the gap. While the off-diagonal terms do decrease exponentially fast $\theta^{2},b^{2}_{x},b^{2}_{y} \sim  1/N$, the diagonal term instead increases as $b^{2}_{z} \sim L$. Putting these scalings into Eq.(~\ref{eq:special-state-energy-gap-Pauli-coefficients}), we see that $\mathcal{E}\left[ \Delta^{2}\right] \sim A/N + B \delta^2 / N + C L \delta^2$ for some order-1 coefficients $A$, $B$, and $C$. Then, for $\delta \gg \delta_{c,\rm comp} \equiv \sqrt{A/(CLN)} \sim 1/\sqrt{L2^L}$, the linear $b_z$ term dominates over the $\theta^{2}$ term giving a power-law-large gap. While this increases the oscillation frequency within the special states, it comes at the cost of exponentially decreasing the oscillation amplitude. In the limit of $\delta \gg \delta_{c,\rm comp}$, the states $|0\rangle$ and $|\Bar{x}\rangle$ become  eigenstates of $G^{\rm spec}(\delta)$ and therefore their probabilities remain at
\begin{eqnarray}
    P_{0}(T) = \frac{1}{N}, \quad P_{\Bar{x}}(T) = 1-\frac{1}{N},
\end{eqnarray}
at all times $T$, up to exponentially small oscillations. Since there is no longer an order-1 probability of finding the target state at any times, this implies that the model has lost computing power. Therefore, our special state analysis unfortunately implies that at this much smaller critical noise $\delta_{c,\rm comp} = \mathcal{O}(2^{-L/2})$ we have a ``computing transition,'' because the algorithm loses computing power for $\delta \apprge \delta_{c,\rm comp}$. By contrast, the gap-closing critical noise $\delta_{c,\rm gap} =  \mathcal{O}(L^{-1})$ is much larger, but occurs well after computational power has been lost.

Given that the large coefficient $b_z$  in $H^{\rm spec}_{\rm eff}$ implies a special basis for the special state manifold, one might hope to recover an exponentially small gap by randomly rotating the basis, e.g., by choosing a different noise model. However, we now argue that this is not possible for any local or $k$-local noise model. The key point is that the relatively larger magnitude of the $b_{z}$ component compared to $b_{x}$ and $b_{y}$ comes from local distinguishability of the states $|\textbf{0}\rangle$ and $|\Bar{x}\rangle$, meaning that generic local (or $k$-local) perturbations in $H_\mathrm{eff}$ will have a (random) diagonal component in this basis. By contrast, distinguishing the off-diagonal cat states $|\textbf{0}\rangle + e^{i\varphi}  |\Bar{x}\rangle$ requires global operators. This implies that $\left| \langle \textbf{0}| H_{\rm eff} | \Bar{x}\rangle \right|^2 \ll \left| \langle  \textbf{0}| H_{\rm eff} | \textbf{0}\rangle\right|^2, \left| \langle \Bar{x} | H_{\rm eff} |\Bar{x}\rangle \right|^2$, which produces $b_z^2 \gg b_{x,y}^2$ for generic $k$-local error models.

\section{Conclusion}

We analyzed Grover’s algorithm with systematic noise by modeling the Grover
operator as a Floquet unitary. We showed that the bulk energy spectrum is well-described
by an effective Hamiltonian whose spectral statistics match the Gaussian unitary ensemble but whose Gaussian eigenvalue distribution reflects the underlying $k$-local nature of the noise model. We use this structure to derive sharp crossovers between behaviors, finding a gap-closing ergodicity transition at relatively large noise strength $\delta_{c,\rm gap} = \mathcal{O}(L^{-1})$ and a separate transition where computational power is lost above an exponentially small noise strength $\delta_{c,\rm comp} = \mathcal{O}(2^{-L/2})$. The random matrix theory model enables analytical predictions for many quantities, such as these noise thresholds and dynamics of the initial state. 

The machinery we have developed for this simplest case of Grover's algorithm with a single solution should provide insights into more generally applying many-body physics approaches to noisy quantum algorithms. For example, we can readily see that Grover's algorithm with multiple solutions exhibits the same phenomenology since the core aspects of a degenerate gapped bulk and cat-like special eigenstates hold there as well. While we specifically focus on unraveling both the oracle and Grover diffusion operator, the same effect will occur if we just unravel the Grover's diffusion operator (reflection about $|x\rangle$) since it is identical to the oracle for a single target state. Introducing noise to only one of these operators, rather than both, would merely adjust the critical disorder thresholds by approximately a factor of 2, affecting only the scaling prefactor rather than the fundamental behavior. A more interesting variant is to instead consider decompositions of the multiqubit Toffoli gate using ancilla qubits. If the ancillas are erased before reuse, they will behave similar to a conventional noise channel which previous results suggest will yield an exponentially small error threshold. On the other hand, if the ancillas are uncomputed and the measured and postselected, the resulting dynamics is noisy but nonunitary, similar to dynamics generated by non-Hermitian Hamiltonians. Non-Hermitian dynamics are known to have qualitatively different phase structures than their Hermitian counterparts. How this affects the phase structure and noise tolerance of Grover's algorithm remains a fascinating open question.

Another important avenue to investigate for Grover's algorithm is the role of non-$k$-local gating. While most controllable quantum computers rely on $k$-local gates - even if swap operations or atomic repositioning relax the assumption of strict spatial locality -- there are potential routes to remove this restriction by using a nonlocal cavity mode or the long-range tails of Rydberg blockade \cite{PRXQuantum.2.020319}. This not only allows more efficient decomposition of the multicontrolled gates, but also implies that systematic gate errors will be nonlocal as well.
Nonlocal errors in $H_\mathrm{eff}$  will cause it to behave more like a proper GUE random matrix, for which a power law noise tolerance $\delta \sim 1/L^\alpha$  can in principle be admitted for both the ergodicity transition and the computational transition. In the NISQ era of finite noise, it is crucial to know these noise scalings in order to determine near-term experimental feasibility. We therefore suggest that nonlocal gating may be a pathway to drastically improved noise tolerance for Grover's algorithm, based on our analytical understanding through random matrix theory. 

In the context of many-body Floquet systems, decorrelating unitary errors between Floquet cycles often has the effect of destabilizing the phase of matter rather than slowing its decay. For example, in the topological Floquet-Anderson insulator,
one of us showed in previous work \cite{PhysRevLett.127.270601}  that it
is stable to weak systematic noise, whereas uncorrelated noise between Floquet cycles leads to eventual decay. A similar situation occurs in many other Floquet
phases of matter, such as time crystals, which are absolutely stable to weak systematic perturbations as one of us showed in \cite{PhysRevB.94.085112} but unstable to nonsystematic noise \cite{PhysRevB.95.195135}.
However, our analysis suggest that the application of twirling operations to decorrelate the noise could significantly mitigate the impact of errors, improving the scaling from $2^{-L/2}$ to $2^{-L/4}$. This observation is consistent with the results of Shapira \emph{et al}. \cite{PhysRevA.67.042301}, although their analysis did not fully consider the effect of unitary noise in the oracle or the Grover diffusion operator.

Finally, while Grover's algorithm provides a concrete example of Floquet analysis in the context of useful quantum algorithms, the eventual goal is to apply similar tools to a wider class of algorithms. Rather than approaching such problem on an individual basis, we would hope to find inspiration for doing so in ``grand unified'' theories of quantum computation \cite{PRXQuantum.2.040203}. While no Floquet structure is immediately apparent in such prescriptions, we might hope to find a constituent block on which similar analysis can be applied and use that insight to understand systematic noise tolerance of a wider class of quantum algorithms.

\section{Acknowledgements}

We thank Sarang Gopalakrishnan, Arjun Mirani, and Junpeng Hou for useful discussions. This work was performed with support from the National Science Foundation (NSF) through Awards No. OMR-2228725 and DMR-1945529, the Welch Foundation through Award No. AT-2036-20200401, and DARPA through Award No. HR00112330022 (M.K. and S.D.). V.K. acknowledges support from the Office of Naval Research Young Investigator Program (ONR YIP) under Award No. N00014-24-1-2098, the Alfred P. Sloan Foundation through a Sloan Research Fellowship and the Packard Foundation through a Packard Fellowship in Science and Engineering. C.Z. acknowledges support from AFOSR (FA9550-20-1-0220), NSF (PHY-2409943, OMR-2228725, ECCS-2411394, OSI-2326628). Part of this work was performed at the Aspen Center for Physics, which is supported by NSF Grant No. PHY-1607611, and at the Kavli Institute for Theoretical Physics, which is supported by NSF Grant No. PHY-1748958.

\appendix

\section{Structure of effective Hamiltonian \label{sec:matrix_elements_of_Heff}}
 
 As shown in the main text, the effective noise Hamiltonian $H_\mathrm{eff}$  does not show full random matrix statistics. Here we analyze the structure further.
 
\begin{figure}
    \centering
{\includegraphics[width=0.48\textwidth]{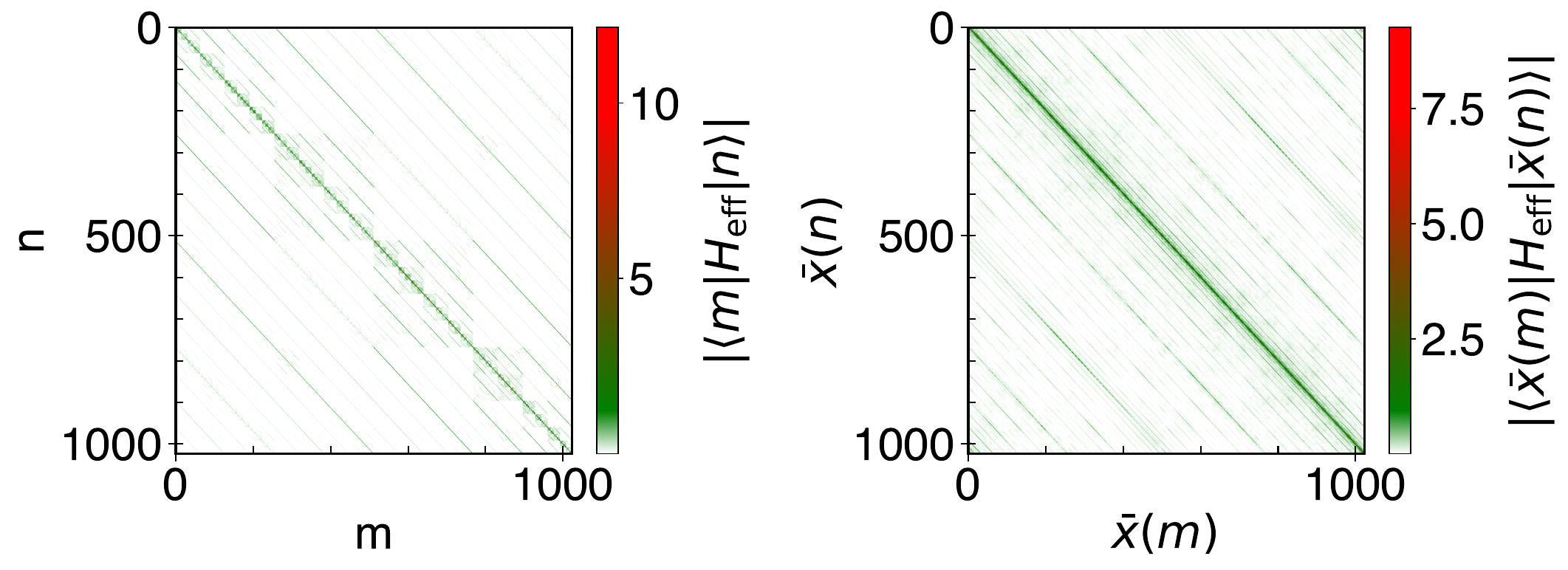}}
    \caption{Heatmaps of the magnitude of the elements of $H_{\rm eff}$ in the computational basis $\{ |k\rangle\}$ (left) and in the eigenbasis of $G_{0}$, $\{ |\Bar{x}(k)\rangle\}$(right) for $L=10$. The colorbar represents the absolute value of the elements in both the bases.}\label{fig:heatmap-h-eff-compt-eigen-basis}
\end{figure}

We start with a heatmap of the magnitude of the elements of $H_{\rm eff}$ in the computational ($z$) basis and in the eigenbasis of $G_{0}$ in Fig \ref{fig:heatmap-h-eff-compt-eigen-basis}. A random matrix would show uniform average color in any basis, with no clear distinction between diagonal and off-diagonal matrix elements. By contrast, $H_\mathrm{eff}$  shows band-like structures with the strongest matrix elements corresponding to the diagonal. This follows from the $k$-locality of the gate perturbations in the structure of $H_{\rm eff}$ (Eq. \eqref{eq:effective-Hamiltonian-definiton}), which means that only a limited number of spins can be flipped in the $z$-basis (and the same applies to the eigenbasis of $G_{0}$).
\section{Dynamics of the special states}

In addition to gap-closing and computational transitions, the random matrix treatment gives analytical predictions for nearly all properties of the algorithm's behavior. In this appendix, we describe the behavior of special state dynamics, which is the most relevant dynamical feature that allows the unperturbed algorithm $G_0$ to be successful. While we can make predictions for how systematic noise modifies these dynamics, their computational relevance is limited to the very small regime of $\delta \ll \delta_{c,\rm comp}$. However, we can also make concrete predictions for ergodicity breaking as long as $\delta < \delta_{c,\rm gap}$.

Before directly examining dynamics, it is worth briefly thinking about the behavior of perturbation theory in this degenerate many-body setting. In particular, given that we start within the special state manifold, does a finite probability of staying in this manifold survive at late time?

In order to address this question, we define the effective Hamiltonian within the special state manifold as
\begin{eqnarray}
    H^{'}(\delta) &=& 2\theta\tau_{y}-\delta H^{\rm spec}_{\rm eff} \nonumber\\
    &=& -\Delta (\cos\theta^{'} \tau_{z}+\sin\theta^{'}(\tau_{x} \cos \phi +\tau_{y} \sin \phi)) \nonumber
\end{eqnarray}
where $\theta^{'} \equiv \theta^{'}(\delta)$. Then we write the eigenstates of $H^{'}(\delta)$ as
\begin{eqnarray}
    |\psi^{(0)}_{1}\rangle &=& \cos\left( \frac{\theta^{'}}{2} \right)| \textbf{0}\rangle+e^{i\phi}\sin\left( \frac{\theta^{'}}{2} \right)|\Bar{x}\rangle,\\
    |\psi^{(0)}_{2}\rangle &=& \sin\left( \frac{\theta^{'}}{2} \right)| \textbf{0}\rangle -e^{i\phi}\cos\left( \frac{\theta^{'}}{2} \right)|\Bar{x}\rangle,
\end{eqnarray}
with energies $\epsilon^{(0)}_{1,2}$ near $\pi$. Then we can write,
\begin{eqnarray}
    G(\delta) &=& e^{-iH_{\rm eff}}G_{0} \approx e^{-i\delta H^{'}_{\rm eff}} e^{-i\delta H^{\rm spec}_{\rm eff}}G_{0} = e^{-i\delta H^{'}_{\rm eff}} \Tilde{G},\nonumber
\end{eqnarray}
where $H_{\rm eff}$ is divided into $H^{\rm spec}_{\rm eff}$, a block containing the two special states, and the remainder $H^{'}_{\rm eff} = H_{\rm eff}-H^{\rm spec}_{\rm eff}$ such that $\Tilde{G}|\psi^{(0)}_{j}\rangle = e^{-i\epsilon^{(0)}_{j}}|\psi^{(0)}_{j}\rangle$. Next, consider the action of $G(\delta)$ on the special states,
\begin{equation}
G(\delta)|\psi_{j}\rangle = e^{-i\delta H^{\rm spec}_{\rm eff}}\Tilde{G} |\psi_{j}\rangle;
\end{equation}
expanding both sides up to the first order in $\delta$, we get
\begin{eqnarray}
    e^{-i\delta H^{'}_{\rm eff}} &\Tilde{G}&(|\psi^{(0)}_{j}\rangle +\delta|\psi^{(1)}_{j}\rangle+\cdots)\nonumber\\
    &=& e^{-i(E^{(0)}_{j}+\delta E^{(1)}_{j}+\cdots)}(|\psi^{(0)}_{j}\rangle+\delta|\psi^{(1)}_{j}\rangle+\cdots).\nonumber
\end{eqnarray}
Taking the matrix element with bulk state $|\Bar{x}_{k}\rangle$ and keeping terms up to order $\delta$, we get
\begin{eqnarray}
\langle \Bar{x}_{k}|\psi^{(1)}_{j}\rangle(1-e^{-iE^{(0)}_{j}}) &=& ie^{-iE^{(0)}_{j}}\langle\Bar{x}_{k}|H^{'}_{\rm eff}|\psi^{(0)}_{j}\rangle,
\end{eqnarray}
where we have used $\langle \Bar{x}_{k}| \Tilde{G}|\psi^{(0)}_{j}\rangle = \langle \Bar{x}_{k}|\psi^{(0)}_{j}\rangle = 0$. Next using $E^{(0)}_{j}\approx \pi$, we obtain
\begin{eqnarray}
    \langle\Bar{x}_{k}|\psi^{(1)}_{j}\rangle &=& \frac{-i}{2}\langle\Bar{x}_{k}|H^{'}_{\rm eff}|\psi^{(0)}_{j}\rangle.
\end{eqnarray}
Therefore, the net bulk occupation in the dressed eigenstate $|\psi_j\rangle$ is approximately 
\begin{eqnarray}
    P_{\text{bulk},j} &=& \sum_{k} |\delta \langle\Bar{x}_{k}|\psi^{(1)}_{j}\rangle|^{2}\nonumber\\
    &=& \frac{\delta^{2}}{4} \sum_{k}|\langle \Bar{x}_{k}|H^{'}_{\rm eff}|\psi^{(0)}_{j}\rangle|^{2}. \label{eq:P_bulk_j}
\end{eqnarray}
While the exact expression for this off-diagonal matrix element, $|\langle \Bar{x}_{k}|H^{'}_{\rm eff}|\psi^{(0)}_{j}\rangle|^{2}$, depends on the particular choice of $k$, the sum over these matrix elements is constrained to be no larger than $E_0^2$  (at least upon averaging over $j$) due to the fact that $\mathrm{Tr}[H_\mathrm{eff}^2] = E_0^2 N$ can be broken up into diagonal and off-diagonal pieces and the diagonals already given the $E_0^2 N$ scaling. Therefore, the off-diagonal sum must give $E_0^2 N$ or smaller, and the lack of summing over $j$  in Eq.(~\ref{eq:P_bulk_j}) removes the multiplicative factor $N$. Assuming that the sum saturates this scaling, we conclude that 
\begin{equation}
     P_{\text{bulk},j} \approx \frac{\delta^{2} E_0^2}{4} \sim \left( \delta / \delta_{c,\rm gap} \right)^2.
\end{equation}
Therefore, the bulk occupation remains finite below $\delta_{c,\rm gap}$, meaning that the special states will also remain finitely occupied at late times and thus break ergodicity as we have claimed.\par
Next, let us analyze the behavior of the probabilities $P_{0}(t)$ and $P_{\Bar{x}}(t)$ in the presence of noise.
Consider the following ansatz for the two special states within first order in noise:
\begin{eqnarray}
|\psi_{1}\rangle &=& \left(1   - P_{b}\right)^{1/4}\left[\cos\left(\frac{\theta^{\prime}}{2}\right)|\bar{x}\rangle+e^{i\phi^{\prime}}\sin\left(\frac{\theta^{\prime}}{2}\right)| \textbf{0}\rangle\right]\nonumber\\
& & + \delta\sum_{k}|\bar{x}_{k}\rangle\langle\bar{x}_{k}|\psi_{1}^{(1)}\rangle,\nonumber\\
|\psi_{2}\rangle &=& \left(1 -P_{b}\right)^{1/4}\left[\sin\left(\frac{\theta^{\prime}}{2}\right)|\bar{x}\rangle-e^{i\phi^{\prime}}\cos\left(\frac{\theta^{\prime}}{2}\right)| \textbf{0}\rangle\right] \nonumber\\
& & + \delta\sum_{k}|\bar{x}_{k}\rangle\langle\bar{x}_{k}|\psi_{2}^{(1)}\rangle,\nonumber
\end{eqnarray}
where we will see that $P_{b}$ is a related parameter that describes late time occupation of the bulk given our actual starting state. This is an unusual ansatz, as it dresses the zeroth order result from degenerate perturbation theory by a nonperturbative overall rescaling of the amplitude, $\left(1   - P_{b}\right)^{1/4}$. This comes from the finite ``leakage'' into the bulk at late times but, as we'll show, seems to give a self-consistent picture of the full dynamics. 

With this ansatz, we can write the initial state of the system as
\begin{eqnarray}
|\psi(0)\rangle &=& |x\rangle\approx|\bar{x}\rangle\nonumber\\
&=& \left(1-P_{b}\right)^{1/4}\left[\cos\left(\frac{\theta^{\prime}}{2}\right)|\psi_{1}\rangle+\sin\left(\frac{\theta^{\prime}}{2}\right)|\psi_{2}\rangle\right]\nonumber\\
& & +\sum_{k}c_{k}|\psi_{k}\rangle,\label{eq:x_bar_eig_decomp}
\end{eqnarray}
where $k$ labels the $N-2$ bulk energy eigenstates. But $|\bar{x}\rangle$ is normalized, so 
\begin{equation}
1=\sqrt{1-P_{b}}+\sum_{k}\left|c_{k}\right|^{2}\implies|c_{k}|^{2}\sim\frac{1-\sqrt{1-P_{b}}}{N}.\nonumber
\end{equation}
Having decomposed the initial state into eigenstates [Eq.(~\ref{eq:x_bar_eig_decomp})],
we can now easily calculate its time evolution as
\begin{eqnarray}
|\psi(t)\rangle &= & \left(1-P_{b}\right)^{1/4} [\cos\left(\frac{\theta^{\prime}}{2}\right)e^{-iE_{1}t}|\psi_{1}\rangle\nonumber\\
& & +\sin\left(\frac{\theta^{\prime}}{2}\right)e^{-iE_{2}t}|\psi_{2}\rangle]
+\sum_{k}c_{k}e^{-iE_{k}t}|\psi_{k}\rangle\nonumber\\
 & & \equiv\sum_{j}c_{j}e^{-iE_{j}t}|\psi_{j}\rangle,\nonumber
\end{eqnarray}
where $t$ is the number of time steps. Then we simply need to calculate observables.

One important observable is the probability to find ourselves in $|\bar{x}\rangle$, which is approximately the Loschmidt echo:
\begin{align}
P_{\bar{x}}(t) & =\left|\langle\bar{x}|\psi(t)\rangle\right|^{2}\approx\left|\langle\psi(0)|\psi(t)\rangle\right|^{2}\nonumber\\
 & =\sum_{jj^{\prime}}\left|c_{j}\right|^{2}\left|c_{j^{\prime}}\right|^{2}e^{-i(E_{j}-E_{j^{\prime}})t}\nonumber\\
 & =\underbrace{P_{\bar{x}}(\text{average as }t\to\infty)}_{\text{A}:~j=j^{\prime}}+\nonumber\\
 & \;\;\;\underbrace{2\sin^{2}\left(\theta^{\prime}\right)\cos^{2}\left(\theta^{\prime}\right)\left(1-P_{b}\right)\cos\left[\left(E_{1}-E_{2}\right)t\right]}_{\text{B: Usual Grover oscillations},~j=1,~j^\prime=2}+\nonumber\\
 & \;\;\;\underbrace{\sum_{k}\left|c_{1}\right|^{2}\left|c_{k}\right|^{2}e^{-i(E_{1}-E_{k})t}+c.c.+\left(1\to2\right)}_{\text{C: Damped oscillations due to bulk},~j=1,~j^\prime=k}+\nonumber\\
 & \;\;\;\underbrace{\sum_{k\neq k^{\prime}}\left|c_{k}\right|^{2}\left|c_{k^{\prime}}\right|^{2}e^{-i(E_{k}-E_{k^{\prime}})t}}_{\text{D: Direct bulk decay},~j=k\neq j^\prime=k^\prime}.\label{eq:Locksmidt-echo-equation}
\end{align}
Let us analyze Eq.\eqref{eq:Locksmidt-echo-equation} term by term.

\underline{Term A: Late-time average}: The existence of a non-exponentially-small
occupation in state $|\bar{x}\rangle$ is a clear indication of nonergodicity.
It will differ between noise realizations, but we can
estimate its noise average (which is equal to the participation ratio
in the eigenstate basis):
\begin{align}
\overline{P_{\bar{x}}} & =\left(1-P_{b}\right)\left[\cos^{4}\left(\frac{\theta^{\prime}}{2}\right)+\sin^{4}\left(\frac{\theta^{\prime}}{2}\right)\right]+\sum_{k}\underbrace{\left|c_{k}\right|^{4}}_{\sim\left(P_{b}/N\right)^{2}}\nonumber\\
 & \approx\left(1-P_{b}\right)\left[\cos^{4}\left(\frac{\theta^{\prime}}{2}\right)+\sin^{4}\left(\frac{\theta^{\prime}}{2}\right)\right]+O(1/N).\nonumber
\end{align}
Before continuing, it's worth doing a similar calculation for the
target state:
\begin{eqnarray}
\overline{P_{0}} &=& \left(\sqrt{1-P_{b}}\right)[\cos^{2}\left(\frac{\theta^{\prime}}{2}\right)\left|\langle0|\psi_{1}\rangle\right|^{2}\nonumber\\
& & +\sin^{2}\left(\frac{\theta^{\prime}}{2}\right)\left|\langle0|\psi_{2}\rangle\right|^{2}]+\underbrace{\sum_{k}\left|c_{k}\right|^{2}\left|\langle0|\psi_{k}\rangle\right|^{2}}_{O(1/N)}\nonumber\\
 & &\approx\left(1-P_{b}\right)\left[2\cos^{2}\left(\frac{\theta^{\prime}}{2}\right)\sin^{2}\left(\frac{\theta^{\prime}}{2}\right)\right]\nonumber\\
\implies\overline{P_{\bar{x}}}+\overline{P_{0}} & & =1-P_{b}\nonumber
\end{eqnarray}
as expected -- $P_b$ is the late-time probability to leave the special state manifold.

\underline{Term B: Grover oscillations}: These are the oscillations
that occur at frequency $\Delta\sim1/\sqrt{N}$ in the absence of noise and with notable modifications in both amplitude and frequency in the range $\delta_{c,\rm comp} < \delta < \delta_{c,\rm gap}$. Specifically, it is clear that, when $P_{b}>0$ and/or
$\theta^{\prime}\neq\pi/2$ (as happens for finite $\delta$), oscillation
amplitude is less than unity.

\underline{Term C: Damped oscillations from bulk continuum}: Due to
the $\pi$ gap between bulk and special states, there are fast (period
2) oscillations in the occupation. However, since the bulk has finite
spectral width, these oscillations are damped. The gap $\Delta$ between $E_{1}$ and $E_{2}$ will not be resolvable
on the decay time scale, so for all intents and purposes we can just
set both of these to $\pi$. Using the Gaussian density of states obtained in the main text, 
\begin{equation}
\nu(E)=\frac{N}{\sqrt{2\pi\delta^{2}E_{0}^{2}}}e^{-E^{2}/(2\delta^{2}E_{0}^{2})},\nonumber
\end{equation}
we get,
\begin{equation}
P_{\bar{x}}^{\text{C}}(t)=2\left(-1\right)^{t}\left(\sqrt{1-P_{b}}-1+P_{b}\right)e^{-\delta^{2}E_{0}^{2}t^{2}/2},\nonumber
\end{equation}
which is simply Gaussian relaxation on the time scale $\tau\sim1/\left(\delta E_{0}\right)\sim\delta_{c,\rm gap}/\delta$
that we expect to see. Notice that this is a fast $O(1)$ time scale
as long we keep $\delta/\delta_{c,\rm gap}$ fixed as we take $L$ large.

\begin{figure*}
\includegraphics[width=0.9\textwidth]{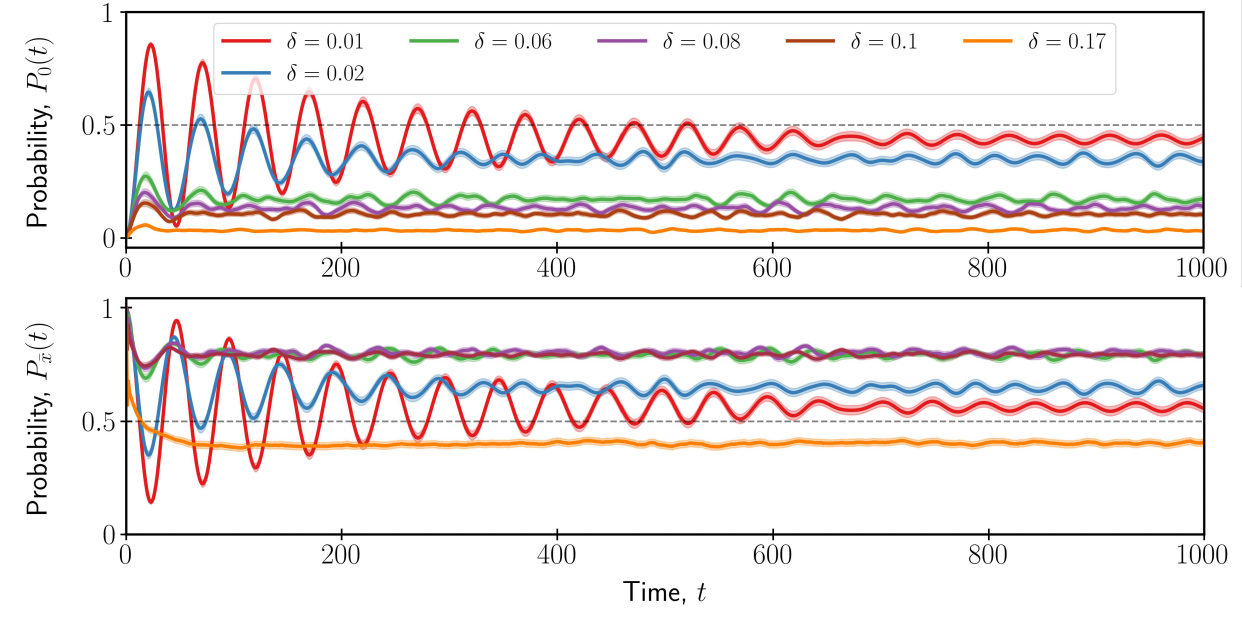}
\caption{Plot of average probability $P_{0}(t)$ and $P_{\Bar{x}}(t)$ for $L=10$ as a function of time steps for different noise strengths: $ 0.01 < \delta_{c,\rm comp} \approx 0.02 < 0.06 < 0.08 < 0.1 < \delta_{c,\rm gap} \approx 0.15 < 0.17$. When $\delta < \delta_{c,\rm gap}$, the probabilities oscillate with a larger frequency but smaller amplitude as compared to the noiseless case. The noises $\delta = 0.06, 0.08,0.1$ lie between $\delta_{c,\rm gap}$ and $\delta_{c,\rm comp}$, for which there is no significant change in the probabilities as anticipated from the above analytical calculation. Noise $\delta = 0.17 > \delta_{c,\rm comp}$ represents the noise dominated dynamics.  Error bands represents the standard error of mean for $256$ noise realizations.}\label{fig:fig:average-of-P_0}
\end{figure*}

\underline{Term D: Bulk decay}: Finally, the last term tells us how Loschmidt echo of a given state -- in this case $|\bar{x}\rangle$
-- decays when propagated by the random bulk matrix $H_{\rm eff}$. Again,
using the Gaussian approximation for the density of states, we find
\begin{eqnarray}
P_{\bar{x}}^{D}(t) &\approx& \frac{\left(1-\sqrt{1-P_{b}}\right)^{2}}{N^{2}}\nonumber\\
&&\times \int dEdE^{\prime}\nu(E)\nu(E^{\prime})\cos\left(\left(E-E^{\prime}\right)t\right),\nonumber\\
 &=&\left(1-\sqrt{1-P_{b}}\right)^{2}e^{-\delta^{2}E_{0}^{2}t^{2}}.\nonumber
\end{eqnarray}
This is similar to the decay rate of the $2T$-periodic oscillations
(term C) with a rate that is $\sqrt{2}$ times faster. Finally, as a sanity check, let us confirm
that $P_{\bar{x}}(0)=1$:
\begin{eqnarray}
P_{\bar{x}}^{A}(0)+p_{\bar{x}}^{B}(0) & = & \left(1-P_{b}\right)\left[\cos^{4}\left(\frac{\theta^{\prime}}{2}\right)+\sin^{4}\left(\frac{\theta^{\prime}}{2}\right)\right],\nonumber\\
& & +2\sin^{2}\left(\theta^{\prime}\right)\cos^{2}\left(\theta^{\prime}\right)\left(1-P_{b}\right),\nonumber\\
 & = & 1-P_{b}.\nonumber\\
P_{\bar{x}}^{C}(0)+P_{\bar{x}}^{D}(0) & = & 2\left(\sqrt{1-P_{b}}-1+P_{b}\right)+\left(1-\sqrt{1-P_{b}}\right)^{2},\nonumber\\
 & = & P_{b}.\nonumber
\end{eqnarray}
Thus
\begin{equation}
    P_{\bar{x}}(0) = 1.\nonumber
\end{equation}
The predicted behaviors above are seen in Fig \ref{fig:fig:average-of-P_0}. Note that the period-$2$ oscillations do appear on close inspection of the short-time behavior (cf. the sharp jump in the curve for $\delta=0.17$  in orange), but are not clear on the longer time scale plotted here.

\bibliography{apssamp}

\providecommand{\noopsort}[1]{}\providecommand{\singleletter}[1]{#1}%
\begin{thebibliography}{42}%
\makeatletter
\providecommand \@ifxundefined [1]{%
 \@ifx{#1\undefined}
}%
\providecommand \@ifnum [1]{%
 \ifnum #1\expandafter \@firstoftwo
 \else \expandafter \@secondoftwo
 \fi
}%
\providecommand \@ifx [1]{%
 \ifx #1\expandafter \@firstoftwo
 \else \expandafter \@secondoftwo
 \fi
}%
\providecommand \natexlab [1]{#1}%
\providecommand \enquote  [1]{``#1''}%
\providecommand \bibnamefont  [1]{#1}%
\providecommand \bibfnamefont [1]{#1}%
\providecommand \citenamefont [1]{#1}%
\providecommand \href@noop [0]{\@secondoftwo}%
\providecommand \href [0]{\begingroup \@sanitize@url \@href}%
\providecommand \@href[1]{\@@startlink{#1}\@@href}%
\providecommand \@@href[1]{\endgroup#1\@@endlink}%
\providecommand \@sanitize@url [0]{\catcode `\\12\catcode `\$12\catcode
  `\&12\catcode `\#12\catcode `\^12\catcode `\_12\catcode `\%12\relax}%
\providecommand \@@startlink[1]{}%
\providecommand \@@endlink[0]{}%
\providecommand \url  [0]{\begingroup\@sanitize@url \@url }%
\providecommand \@url [1]{\endgroup\@href {#1}{\urlprefix }}%
\providecommand \urlprefix  [0]{URL }%
\providecommand \Eprint [0]{\href }%
\providecommand \doibase [0]{https://doi.org/}%
\providecommand \selectlanguage [0]{\@gobble}%
\providecommand \bibinfo  [0]{\@secondoftwo}%
\providecommand \bibfield  [0]{\@secondoftwo}%
\providecommand \translation [1]{[#1]}%
\providecommand \BibitemOpen [0]{}%
\providecommand \bibitemStop [0]{}%
\providecommand \bibitemNoStop [0]{.\EOS\space}%
\providecommand \EOS [0]{\spacefactor3000\relax}%
\providecommand \BibitemShut  [1]{\csname bibitem#1\endcsname}%
\let\auto@bib@innerbib\@empty
\bibitem [{\citenamefont {Preskill}(2018)}]{Preskill2018quantumcomputingin}%
  \BibitemOpen
  \bibfield  {author} {\bibinfo {author} {\bibfnamefont {J.}~\bibnamefont
  {Preskill}},\ }\bibfield  {title} {\bibinfo {title} {Quantum {C}omputing in
  the {NISQ} era and beyond},\ }\href
  {https://doi.org/10.22331/q-2018-08-06-79} {\bibfield  {journal} {\bibinfo
  {journal} {{Quantum}}\ }\textbf {\bibinfo {volume} {2}},\ \bibinfo {pages}
  {79} (\bibinfo {year} {2018})}\BibitemShut {NoStop}%
\bibitem [{\citenamefont {Fauseweh}(2024)}]{Fauseweh2024}%
  \BibitemOpen
  \bibfield  {author} {\bibinfo {author} {\bibfnamefont {B.}~\bibnamefont
  {Fauseweh}},\ }\bibfield  {title} {\bibinfo {title} {Quantum many-body
  simulations on digital quantum computers: State-of-the-art and future
  challenges},\ }\href {https://doi.org/10.1038/s41467-024-46402-9} {\bibfield
  {journal} {\bibinfo  {journal} {Nature Communications}\ }\textbf {\bibinfo
  {volume} {15}},\ \bibinfo {pages} {2123} (\bibinfo {year}
  {2024})}\BibitemShut {NoStop}%
\bibitem [{\citenamefont {Bharti}\ \emph {et~al.}(2022)\citenamefont {Bharti},
  \citenamefont {Cervera-Lierta}, \citenamefont {Kyaw}, \citenamefont {Haug},
  \citenamefont {Alperin-Lea}, \citenamefont {Anand}, \citenamefont {Degroote},
  \citenamefont {Heimonen}, \citenamefont {Kottmann}, \citenamefont {Menke},
  \citenamefont {Mok}, \citenamefont {Sim}, \citenamefont {Kwek},\ and\
  \citenamefont {Aspuru-Guzik}}]{RevModPhys.94.015004}%
  \BibitemOpen
  \bibfield  {author} {\bibinfo {author} {\bibfnamefont {K.}~\bibnamefont
  {Bharti}}, \bibinfo {author} {\bibfnamefont {A.}~\bibnamefont
  {Cervera-Lierta}}, \bibinfo {author} {\bibfnamefont {T.~H.}\ \bibnamefont
  {Kyaw}}, \bibinfo {author} {\bibfnamefont {T.}~\bibnamefont {Haug}}, \bibinfo
  {author} {\bibfnamefont {S.}~\bibnamefont {Alperin-Lea}}, \bibinfo {author}
  {\bibfnamefont {A.}~\bibnamefont {Anand}}, \bibinfo {author} {\bibfnamefont
  {M.}~\bibnamefont {Degroote}}, \bibinfo {author} {\bibfnamefont
  {H.}~\bibnamefont {Heimonen}}, \bibinfo {author} {\bibfnamefont {J.~S.}\
  \bibnamefont {Kottmann}}, \bibinfo {author} {\bibfnamefont {T.}~\bibnamefont
  {Menke}}, \bibinfo {author} {\bibfnamefont {W.-K.}\ \bibnamefont {Mok}},
  \bibinfo {author} {\bibfnamefont {S.}~\bibnamefont {Sim}}, \bibinfo {author}
  {\bibfnamefont {L.-C.}\ \bibnamefont {Kwek}},\ and\ \bibinfo {author}
  {\bibfnamefont {A.}~\bibnamefont {Aspuru-Guzik}},\ }\bibfield  {title}
  {\bibinfo {title} {Noisy intermediate-scale quantum algorithms},\ }\href
  {https://doi.org/10.1103/RevModPhys.94.015004} {\bibfield  {journal}
  {\bibinfo  {journal} {Rev. Mod. Phys.}\ }\textbf {\bibinfo {volume} {94}},\
  \bibinfo {pages} {015004} (\bibinfo {year} {2022})}\BibitemShut {NoStop}%
\bibitem [{\citenamefont {Grover}(1996)}]{10.1145/237814.237866}%
  \BibitemOpen
  \bibfield  {author} {\bibinfo {author} {\bibfnamefont {L.~K.}\ \bibnamefont
  {Grover}},\ }\bibfield  {title} {\bibinfo {title} {A fast quantum mechanical
  algorithm for database search},\ }in\ \href
  {https://doi.org/10.1145/237814.237866} {\emph {\bibinfo {booktitle}
  {Proceedings of the Twenty-Eighth Annual ACM Symposium on Theory of
  Computing}}},\ \bibinfo {series and number} {STOC '96}\ (\bibinfo
  {publisher} {Association for Computing Machinery},\ \bibinfo {address} {New
  York, NY, USA},\ \bibinfo {year} {1996})\ p.\ \bibinfo {pages}
  {212–219}\BibitemShut {NoStop}%
\bibitem [{\citenamefont {Figgatt}\ \emph {et~al.}(2017)\citenamefont
  {Figgatt}, \citenamefont {Maslov}, \citenamefont {Landsman}, \citenamefont
  {Linke}, \citenamefont {Debnath},\ and\ \citenamefont
  {Monroe}}]{Figgatt2017Complete3G}%
  \BibitemOpen
  \bibfield  {author} {\bibinfo {author} {\bibfnamefont {C.}~\bibnamefont
  {Figgatt}}, \bibinfo {author} {\bibfnamefont {D.~L.}\ \bibnamefont {Maslov}},
  \bibinfo {author} {\bibfnamefont {K.~A.}\ \bibnamefont {Landsman}}, \bibinfo
  {author} {\bibfnamefont {N.~M.}\ \bibnamefont {Linke}}, \bibinfo {author}
  {\bibfnamefont {S.}~\bibnamefont {Debnath}},\ and\ \bibinfo {author}
  {\bibfnamefont {C.~R.}\ \bibnamefont {Monroe}},\ }\bibfield  {title}
  {\bibinfo {title} {Complete 3-qubit grover search on a programmable quantum
  computer},\ }\href@noop {} {\bibfield  {journal} {\bibinfo  {journal} {Nature
  Communications}\ }\textbf {\bibinfo {volume} {8}} (\bibinfo {year}
  {2017})}\BibitemShut {NoStop}%
\bibitem [{\citenamefont {Zhang}\ \emph {et~al.}(2021)\citenamefont {Zhang},
  \citenamefont {Rao}, \citenamefont {Yu}, \citenamefont {Lim},\ and\
  \citenamefont {Korepin}}]{Zhang2021ImplementationOE}%
  \BibitemOpen
  \bibfield  {author} {\bibinfo {author} {\bibfnamefont {K.}~\bibnamefont
  {Zhang}}, \bibinfo {author} {\bibfnamefont {P.}~\bibnamefont {Rao}}, \bibinfo
  {author} {\bibfnamefont {K.}~\bibnamefont {Yu}}, \bibinfo {author}
  {\bibfnamefont {H.}~\bibnamefont {Lim}},\ and\ \bibinfo {author}
  {\bibfnamefont {V.~E.}\ \bibnamefont {Korepin}},\ }\bibfield  {title}
  {\bibinfo {title} {Implementation of efficient quantum search algorithms on
  nisq computers},\ }\href@noop {} {\bibfield  {journal} {\bibinfo  {journal}
  {Quantum Information Processing}\ }\textbf {\bibinfo {volume} {20}} (\bibinfo
  {year} {2021})}\BibitemShut {NoStop}%
\bibitem [{\citenamefont {Zhang}\ \emph {et~al.}(2022)\citenamefont {Zhang},
  \citenamefont {Yu},\ and\ \citenamefont {Korepin}}]{Zhang2022QuantumSO}%
  \BibitemOpen
  \bibfield  {author} {\bibinfo {author} {\bibfnamefont {K.}~\bibnamefont
  {Zhang}}, \bibinfo {author} {\bibfnamefont {K.}~\bibnamefont {Yu}},\ and\
  \bibinfo {author} {\bibfnamefont {V.~E.}\ \bibnamefont {Korepin}},\
  }\bibfield  {title} {\bibinfo {title} {Quantum search on noisy
  intermediate-scale quantum devices},\ }\href@noop {} {\bibfield  {journal}
  {\bibinfo  {journal} {Europhysics Letters}\ }\textbf {\bibinfo {volume}
  {140}} (\bibinfo {year} {2022})}\BibitemShut {NoStop}%
\bibitem [{\citenamefont {Long}\ \emph {et~al.}(2000)\citenamefont {Long},
  \citenamefont {Li}, \citenamefont {Zhang},\ and\ \citenamefont
  {Tu}}]{PhysRevA.61.042305}%
  \BibitemOpen
  \bibfield  {author} {\bibinfo {author} {\bibfnamefont {G.~L.}\ \bibnamefont
  {Long}}, \bibinfo {author} {\bibfnamefont {Y.~S.}\ \bibnamefont {Li}},
  \bibinfo {author} {\bibfnamefont {W.~L.}\ \bibnamefont {Zhang}},\ and\
  \bibinfo {author} {\bibfnamefont {C.~C.}\ \bibnamefont {Tu}},\ }\bibfield
  {title} {\bibinfo {title} {Dominant gate imperfection in grover's quantum
  search algorithm},\ }\href {https://doi.org/10.1103/PhysRevA.61.042305}
  {\bibfield  {journal} {\bibinfo  {journal} {Phys. Rev. A}\ }\textbf {\bibinfo
  {volume} {61}},\ \bibinfo {pages} {042305} (\bibinfo {year}
  {2000})}\BibitemShut {NoStop}%
\bibitem [{\citenamefont {Shenvi}\ \emph {et~al.}(2003)\citenamefont {Shenvi},
  \citenamefont {Brown},\ and\ \citenamefont {Whaley}}]{PhysRevA.68.052313}%
  \BibitemOpen
  \bibfield  {author} {\bibinfo {author} {\bibfnamefont {N.}~\bibnamefont
  {Shenvi}}, \bibinfo {author} {\bibfnamefont {K.~R.}\ \bibnamefont {Brown}},\
  and\ \bibinfo {author} {\bibfnamefont {K.~B.}\ \bibnamefont {Whaley}},\
  }\bibfield  {title} {\bibinfo {title} {Effects of a random noisy oracle on
  search algorithm complexity},\ }\href
  {https://doi.org/10.1103/PhysRevA.68.052313} {\bibfield  {journal} {\bibinfo
  {journal} {Phys. Rev. A}\ }\textbf {\bibinfo {volume} {68}},\ \bibinfo
  {pages} {052313} (\bibinfo {year} {2003})}\BibitemShut {NoStop}%
\bibitem [{\citenamefont {Salas}(2008)}]{Salas2008NoiseEO}%
  \BibitemOpen
  \bibfield  {author} {\bibinfo {author} {\bibfnamefont {P.~J.}\ \bibnamefont
  {Salas}},\ }\bibfield  {title} {\bibinfo {title} {Noise effect on grover
  algorithm},\ }\href@noop {} {\bibfield  {journal} {\bibinfo  {journal} {The
  European Physical Journal D}\ }\textbf {\bibinfo {volume} {46}},\ \bibinfo
  {pages} {365} (\bibinfo {year} {2008})}\BibitemShut {NoStop}%
\bibitem [{\citenamefont {Pablo-Norman}\ and\ \citenamefont
  {Ruiz-Altaba}(1999)}]{PhysRevA.61.012301}%
  \BibitemOpen
  \bibfield  {author} {\bibinfo {author} {\bibfnamefont {B.}~\bibnamefont
  {Pablo-Norman}}\ and\ \bibinfo {author} {\bibfnamefont {M.}~\bibnamefont
  {Ruiz-Altaba}},\ }\bibfield  {title} {\bibinfo {title} {Noise in grover's
  quantum search algorithm},\ }\href
  {https://doi.org/10.1103/PhysRevA.61.012301} {\bibfield  {journal} {\bibinfo
  {journal} {Phys. Rev. A}\ }\textbf {\bibinfo {volume} {61}},\ \bibinfo
  {pages} {012301} (\bibinfo {year} {1999})}\BibitemShut {NoStop}%
\bibitem [{\citenamefont {Shapira}\ \emph {et~al.}(2003)\citenamefont
  {Shapira}, \citenamefont {Mozes},\ and\ \citenamefont
  {Biham}}]{PhysRevA.67.042301}%
  \BibitemOpen
  \bibfield  {author} {\bibinfo {author} {\bibfnamefont {D.}~\bibnamefont
  {Shapira}}, \bibinfo {author} {\bibfnamefont {S.}~\bibnamefont {Mozes}},\
  and\ \bibinfo {author} {\bibfnamefont {O.}~\bibnamefont {Biham}},\ }\bibfield
   {title} {\bibinfo {title} {Effect of unitary noise on grover's quantum
  search algorithm},\ }\href {https://doi.org/10.1103/PhysRevA.67.042301}
  {\bibfield  {journal} {\bibinfo  {journal} {Phys. Rev. A}\ }\textbf {\bibinfo
  {volume} {67}},\ \bibinfo {pages} {042301} (\bibinfo {year}
  {2003})}\BibitemShut {NoStop}%
\bibitem [{\citenamefont {Chen}\ \emph {et~al.}(2023)\citenamefont {Chen},
  \citenamefont {Cotler}, \citenamefont {Huang},\ and\ \citenamefont
  {Li}}]{Chen2023}%
  \BibitemOpen
  \bibfield  {author} {\bibinfo {author} {\bibfnamefont {S.}~\bibnamefont
  {Chen}}, \bibinfo {author} {\bibfnamefont {J.}~\bibnamefont {Cotler}},
  \bibinfo {author} {\bibfnamefont {H.-Y.}\ \bibnamefont {Huang}},\ and\
  \bibinfo {author} {\bibfnamefont {J.}~\bibnamefont {Li}},\ }\bibfield
  {title} {\bibinfo {title} {The complexity of nisq},\ }\href
  {https://doi.org/10.1038/s41467-023-41217-6} {\bibfield  {journal} {\bibinfo
  {journal} {Nature Communications}\ }\textbf {\bibinfo {volume} {14}},\
  \bibinfo {pages} {6001} (\bibinfo {year} {2023})}\BibitemShut {NoStop}%
\bibitem [{\citenamefont {Boyer}\ \emph {et~al.}(1998)\citenamefont {Boyer},
  \citenamefont {Brassard}, \citenamefont {Høyer},\ and\ \citenamefont
  {Tapp}}]{Boyer1996}%
  \BibitemOpen
  \bibfield  {author} {\bibinfo {author} {\bibfnamefont {M.}~\bibnamefont
  {Boyer}}, \bibinfo {author} {\bibfnamefont {G.}~\bibnamefont {Brassard}},
  \bibinfo {author} {\bibfnamefont {P.}~\bibnamefont {Høyer}},\ and\ \bibinfo
  {author} {\bibfnamefont {A.}~\bibnamefont {Tapp}},\ }\bibfield  {title}
  {\bibinfo {title} {Tight bounds on quantum searching},\ }\href
  {https://doi.org/https://doi.org/10.1002/(SICI)1521-3978(199806)46:4/5<493::AID-PROP493>3.0.CO;2-P}
  {\bibfield  {journal} {\bibinfo  {journal} {Fortschritte der Physik}\
  }\textbf {\bibinfo {volume} {46}},\ \bibinfo {pages} {493} (\bibinfo {year}
  {1998})}\BibitemShut {NoStop}%
\bibitem [{\citenamefont {Zalka}(1999)}]{PhysRevA.60.2746}%
  \BibitemOpen
  \bibfield  {author} {\bibinfo {author} {\bibfnamefont {C.}~\bibnamefont
  {Zalka}},\ }\bibfield  {title} {\bibinfo {title} {Grover's quantum searching
  algorithm is optimal},\ }\href {https://doi.org/10.1103/PhysRevA.60.2746}
  {\bibfield  {journal} {\bibinfo  {journal} {Phys. Rev. A}\ }\textbf {\bibinfo
  {volume} {60}},\ \bibinfo {pages} {2746} (\bibinfo {year}
  {1999})}\BibitemShut {NoStop}%
\bibitem [{\citenamefont {Oka}\ and\ \citenamefont
  {Kitamura}(2019)}]{OkaReview2019}%
  \BibitemOpen
  \bibfield  {author} {\bibinfo {author} {\bibfnamefont {T.}~\bibnamefont
  {Oka}}\ and\ \bibinfo {author} {\bibfnamefont {S.}~\bibnamefont {Kitamura}},\
  }\bibfield  {title} {\bibinfo {title} {Floquet engineering of quantum
  materials},\ }\href
  {https://doi.org/10.1146/annurev-conmatphys-031218-013423} {\bibfield
  {journal} {\bibinfo  {journal} {Annual Review of Condensed Matter Physics}\
  }\textbf {\bibinfo {volume} {10}},\ \bibinfo {pages} {387} (\bibinfo {year}
  {2019})}\BibitemShut {NoStop}%
\bibitem [{\citenamefont {Bukov}\ \emph {et~al.}(2015)\citenamefont {Bukov},
  \citenamefont {D'Alessio},\ and\ \citenamefont
  {Polkovnikov}}]{doi:10.1080/00018732.2015.1055918}%
  \BibitemOpen
  \bibfield  {author} {\bibinfo {author} {\bibfnamefont {M.}~\bibnamefont
  {Bukov}}, \bibinfo {author} {\bibfnamefont {L.}~\bibnamefont {D'Alessio}},\
  and\ \bibinfo {author} {\bibfnamefont {A.}~\bibnamefont {Polkovnikov}},\
  }\bibfield  {title} {\bibinfo {title} {Universal high-frequency behavior of
  periodically driven systems: from dynamical stabilization to floquet
  engineering},\ }\href {https://doi.org/10.1080/00018732.2015.1055918}
  {\bibfield  {journal} {\bibinfo  {journal} {Advances in Physics}\ }\textbf
  {\bibinfo {volume} {64}},\ \bibinfo {pages} {139} (\bibinfo {year}
  {2015})}\BibitemShut {NoStop}%
\bibitem [{\citenamefont {Floquet}()}]{FloquetSurL}%
  \BibitemOpen
  \bibfield  {author} {\bibinfo {author} {\bibfnamefont {G.}~\bibnamefont
  {Floquet}},\ }\bibfield  {title} {\bibinfo {title} {Sur les {\'e}quations
  diff{\'e}rentielles lin{\'e}aires {\`a} coefficients p{\'e}riodiques},\
  }\href@noop {} {\bibfield  {journal} {\bibinfo  {journal} {Annales
  Scientifiques De L Ecole Normale Superieure}\ }\textbf {\bibinfo {volume}
  {12}},\ \bibinfo {pages} {47}}\BibitemShut {NoStop}%
\bibitem [{\citenamefont {Nielsen}\ and\ \citenamefont
  {Chuang}(2010)}]{nielsen_chuang_2010}%
  \BibitemOpen
  \bibfield  {author} {\bibinfo {author} {\bibfnamefont {M.~A.}\ \bibnamefont
  {Nielsen}}\ and\ \bibinfo {author} {\bibfnamefont {I.~L.}\ \bibnamefont
  {Chuang}},\ }\href@noop {} {\emph {\bibinfo {title} {Quantum Computation and
  Quantum Information: 10th Anniversary Edition}}}\ (\bibinfo  {publisher}
  {Cambridge University Press},\ \bibinfo {year} {2010})\BibitemShut {NoStop}%
\bibitem [{\citenamefont {Barenco}\ \emph {et~al.}(1995)\citenamefont
  {Barenco}, \citenamefont {Bennett}, \citenamefont {Cleve}, \citenamefont
  {DiVincenzo}, \citenamefont {Margolus}, \citenamefont {Shor}, \citenamefont
  {Sleator}, \citenamefont {Smolin},\ and\ \citenamefont
  {Weinfurter}}]{PhysRevA.52.3457}%
  \BibitemOpen
  \bibfield  {author} {\bibinfo {author} {\bibfnamefont {A.}~\bibnamefont
  {Barenco}}, \bibinfo {author} {\bibfnamefont {C.~H.}\ \bibnamefont
  {Bennett}}, \bibinfo {author} {\bibfnamefont {R.}~\bibnamefont {Cleve}},
  \bibinfo {author} {\bibfnamefont {D.~P.}\ \bibnamefont {DiVincenzo}},
  \bibinfo {author} {\bibfnamefont {N.}~\bibnamefont {Margolus}}, \bibinfo
  {author} {\bibfnamefont {P.}~\bibnamefont {Shor}}, \bibinfo {author}
  {\bibfnamefont {T.}~\bibnamefont {Sleator}}, \bibinfo {author} {\bibfnamefont
  {J.~A.}\ \bibnamefont {Smolin}},\ and\ \bibinfo {author} {\bibfnamefont
  {H.}~\bibnamefont {Weinfurter}},\ }\bibfield  {title} {\bibinfo {title}
  {Elementary gates for quantum computation},\ }\href
  {https://doi.org/10.1103/PhysRevA.52.3457} {\bibfield  {journal} {\bibinfo
  {journal} {Phys. Rev. A}\ }\textbf {\bibinfo {volume} {52}},\ \bibinfo
  {pages} {3457} (\bibinfo {year} {1995})}\BibitemShut {NoStop}%
\bibitem [{\citenamefont {Saeedi}\ and\ \citenamefont
  {Pedram}(2013)}]{PhysRevA.87.062318}%
  \BibitemOpen
  \bibfield  {author} {\bibinfo {author} {\bibfnamefont {M.}~\bibnamefont
  {Saeedi}}\ and\ \bibinfo {author} {\bibfnamefont {M.}~\bibnamefont
  {Pedram}},\ }\bibfield  {title} {\bibinfo {title} {Linear-depth quantum
  circuits for $n$-qubit toffoli gates with no ancilla},\ }\href
  {https://doi.org/10.1103/PhysRevA.87.062318} {\bibfield  {journal} {\bibinfo
  {journal} {Phys. Rev. A}\ }\textbf {\bibinfo {volume} {87}},\ \bibinfo
  {pages} {062318} (\bibinfo {year} {2013})}\BibitemShut {NoStop}%
\bibitem [{Note1()}]{Note1}%
  \BibitemOpen
  \bibinfo {note} {For a rigorous argument for the linear depth of this
  decomposition, see \cite {PhysRevA.87.062318,
  PhysRevA.106.042602}}\BibitemShut {NoStop}%
\bibitem [{Note2()}]{Note2}%
  \BibitemOpen
  \bibinfo {note} {The decomposition shown requires gate range proportional to
  $L$, but the authors argue that finite-range gates still allow depth
  $O(L)$.}\BibitemShut {Stop}%
\bibitem [{\citenamefont {Page}(1993)}]{PhysRevLett.71.1291}%
  \BibitemOpen
  \bibfield  {author} {\bibinfo {author} {\bibfnamefont {D.~N.}\ \bibnamefont
  {Page}},\ }\bibfield  {title} {\bibinfo {title} {Average entropy of a
  subsystem},\ }\href {https://doi.org/10.1103/PhysRevLett.71.1291} {\bibfield
  {journal} {\bibinfo  {journal} {Phys. Rev. Lett.}\ }\textbf {\bibinfo
  {volume} {71}},\ \bibinfo {pages} {1291} (\bibinfo {year}
  {1993})}\BibitemShut {NoStop}%
\bibitem [{Note3()}]{Note3}%
  \BibitemOpen
  \bibinfo {note} {Here, the origin of time in the Heisenberg picture is
  defined as the end of the Floquet cycle, such that Heisenberg evolution
  ``rewinds'' the action of $e^{-iH_M \theta _M} e^{-iH_{M-1} \theta _{M-1}}
  \protect \cdots $}\BibitemShut {NoStop}%
\bibitem [{\citenamefont {Oganesyan}\ and\ \citenamefont
  {Huse}(2007)}]{PhysRevB.75.155111}%
  \BibitemOpen
  \bibfield  {author} {\bibinfo {author} {\bibfnamefont {V.}~\bibnamefont
  {Oganesyan}}\ and\ \bibinfo {author} {\bibfnamefont {D.~A.}\ \bibnamefont
  {Huse}},\ }\bibfield  {title} {\bibinfo {title} {Localization of interacting
  fermions at high temperature},\ }\href
  {https://doi.org/10.1103/PhysRevB.75.155111} {\bibfield  {journal} {\bibinfo
  {journal} {Phys. Rev. B}\ }\textbf {\bibinfo {volume} {75}},\ \bibinfo
  {pages} {155111} (\bibinfo {year} {2007})}\BibitemShut {NoStop}%
\bibitem [{\citenamefont {Kullback}\ and\ \citenamefont
  {Leibler}(1951)}]{10.1214/aoms/1177729694}%
  \BibitemOpen
  \bibfield  {author} {\bibinfo {author} {\bibfnamefont {S.}~\bibnamefont
  {Kullback}}\ and\ \bibinfo {author} {\bibfnamefont {R.~A.}\ \bibnamefont
  {Leibler}},\ }\bibfield  {title} {\bibinfo {title} {{On Information and
  Sufficiency}},\ }\href {https://doi.org/10.1214/aoms/1177729694} {\bibfield
  {journal} {\bibinfo  {journal} {The Annals of Mathematical Statistics}\
  }\textbf {\bibinfo {volume} {22}},\ \bibinfo {pages} {79 } (\bibinfo {year}
  {1951})}\BibitemShut {NoStop}%
\bibitem [{\citenamefont {Luitz}\ \emph {et~al.}(2015)\citenamefont {Luitz},
  \citenamefont {Laflorencie},\ and\ \citenamefont
  {Alet}}]{PhysRevB.91.081103}%
  \BibitemOpen
  \bibfield  {author} {\bibinfo {author} {\bibfnamefont {D.~J.}\ \bibnamefont
  {Luitz}}, \bibinfo {author} {\bibfnamefont {N.}~\bibnamefont {Laflorencie}},\
  and\ \bibinfo {author} {\bibfnamefont {F.}~\bibnamefont {Alet}},\ }\bibfield
  {title} {\bibinfo {title} {Many-body localization edge in the random-field
  heisenberg chain},\ }\href {https://doi.org/10.1103/PhysRevB.91.081103}
  {\bibfield  {journal} {\bibinfo  {journal} {Phys. Rev. B}\ }\textbf {\bibinfo
  {volume} {91}},\ \bibinfo {pages} {081103} (\bibinfo {year}
  {2015})}\BibitemShut {NoStop}%
\bibitem [{\citenamefont {Atas}\ \emph {et~al.}(2013)\citenamefont {Atas},
  \citenamefont {Bogomolny}, \citenamefont {Giraud},\ and\ \citenamefont
  {Roux}}]{PhysRevLett.110.084101}%
  \BibitemOpen
  \bibfield  {author} {\bibinfo {author} {\bibfnamefont {Y.~Y.}\ \bibnamefont
  {Atas}}, \bibinfo {author} {\bibfnamefont {E.}~\bibnamefont {Bogomolny}},
  \bibinfo {author} {\bibfnamefont {O.}~\bibnamefont {Giraud}},\ and\ \bibinfo
  {author} {\bibfnamefont {G.}~\bibnamefont {Roux}},\ }\bibfield  {title}
  {\bibinfo {title} {Distribution of the ratio of consecutive level spacings in
  random matrix ensembles},\ }\href
  {https://doi.org/10.1103/PhysRevLett.110.084101} {\bibfield  {journal}
  {\bibinfo  {journal} {Phys. Rev. Lett.}\ }\textbf {\bibinfo {volume} {110}},\
  \bibinfo {pages} {084101} (\bibinfo {year} {2013})}\BibitemShut {NoStop}%
\bibitem [{Note4()}]{Note4}%
  \BibitemOpen
  \bibinfo {note} {We calculated the KL divergence of the standard Gaussian
  Unitary Ensemble (GUE) by generating GUE matrices of size $2^{14}\times
  2^{14}$ using the package TeNPy}\BibitemShut {NoStop}%
\bibitem [{\citenamefont {Rodriguez-Nieva}\ \emph {et~al.}(2023)\citenamefont
  {Rodriguez-Nieva}, \citenamefont {Jonay},\ and\ \citenamefont
  {Khemani}}]{rodriguez2023quantifying}%
  \BibitemOpen
  \bibfield  {author} {\bibinfo {author} {\bibfnamefont {J.~F.}\ \bibnamefont
  {Rodriguez-Nieva}}, \bibinfo {author} {\bibfnamefont {C.}~\bibnamefont
  {Jonay}},\ and\ \bibinfo {author} {\bibfnamefont {V.}~\bibnamefont
  {Khemani}},\ }\bibfield  {title} {\bibinfo {title} {Quantifying quantum chaos
  through microcanonical distributions of entanglement},\ }\href@noop {}
  {\bibfield  {journal} {\bibinfo  {journal} {arXiv preprint arXiv:2305.11940}\
  } (\bibinfo {year} {2023})}\BibitemShut {NoStop}%
\bibitem [{Note5()}]{Note5}%
  \BibitemOpen
  \bibinfo {note} {The trace is explicitly given by $\protect \frac {\protect
  \text {tr}(H_{\protect \rm eff})}{N} = \DOTSB \sum@ \slimits@ _{k\in
  \{CR_{x}\}}\protect \frac {\epsilon _{k}}{4}+\DOTSB \sum@ \slimits@ _{l\in
  \{Z,H,X\}}\epsilon _{l}$, whose noise-averaged value is equal to zero, while
  its standard deviation is extensive: $\protect \text {std}\left [\protect
  \frac {\protect \text {tr}(H_{\protect \rm eff})}{N} \right ] \sim L$ given
  $O(L^2)$ gates}\BibitemShut {NoStop}%
\bibitem [{\citenamefont {Gopalakrishnan}\ and\ \citenamefont
  {Huse}(2019)}]{PhysRevB.99.134305}%
  \BibitemOpen
  \bibfield  {author} {\bibinfo {author} {\bibfnamefont {S.}~\bibnamefont
  {Gopalakrishnan}}\ and\ \bibinfo {author} {\bibfnamefont {D.~A.}\
  \bibnamefont {Huse}},\ }\bibfield  {title} {\bibinfo {title} {Instability of
  many-body localized systems as a phase transition in a nonstandard
  thermodynamic limit},\ }\href {https://doi.org/10.1103/PhysRevB.99.134305}
  {\bibfield  {journal} {\bibinfo  {journal} {Phys. Rev. B}\ }\textbf {\bibinfo
  {volume} {99}},\ \bibinfo {pages} {134305} (\bibinfo {year}
  {2019})}\BibitemShut {NoStop}%
\bibitem [{\citenamefont {Haake}(2010)}]{Haake_2010}%
  \BibitemOpen
  \bibfield  {author} {\bibinfo {author} {\bibfnamefont {F.}~\bibnamefont
  {Haake}},\ }\href@noop {} {\emph {\bibinfo {title} {Quantum Signatures of
  Chaos}}}\ (\bibinfo  {publisher} {Springer Berlin},\ \bibinfo {year}
  {2010})\BibitemShut {NoStop}%
\bibitem [{\citenamefont {Bohigas}\ and\ \citenamefont
  {Flores}(1971)}]{BOHIGAS1971261}%
  \BibitemOpen
  \bibfield  {author} {\bibinfo {author} {\bibfnamefont {O.}~\bibnamefont
  {Bohigas}}\ and\ \bibinfo {author} {\bibfnamefont {J.}~\bibnamefont
  {Flores}},\ }\bibfield  {title} {\bibinfo {title} {Two-body random
  hamiltonian and level density},\ }\href
  {https://doi.org/https://doi.org/10.1016/0370-2693(71)90598-3} {\bibfield
  {journal} {\bibinfo  {journal} {Physics Letters B}\ }\textbf {\bibinfo
  {volume} {34}},\ \bibinfo {pages} {261} (\bibinfo {year} {1971})}\BibitemShut
  {NoStop}%
\bibitem [{\citenamefont {French}\ and\ \citenamefont
  {Wong}(1970)}]{FRENCH1970449}%
  \BibitemOpen
  \bibfield  {author} {\bibinfo {author} {\bibfnamefont {J.}~\bibnamefont
  {French}}\ and\ \bibinfo {author} {\bibfnamefont {S.}~\bibnamefont {Wong}},\
  }\bibfield  {title} {\bibinfo {title} {Validity of random matrix theories for
  many-particle systems},\ }\href
  {https://doi.org/https://doi.org/10.1016/0370-2693(70)90213-3} {\bibfield
  {journal} {\bibinfo  {journal} {Physics Letters B}\ }\textbf {\bibinfo
  {volume} {33}},\ \bibinfo {pages} {449} (\bibinfo {year} {1970})}\BibitemShut
  {NoStop}%
\bibitem [{\citenamefont {Torres-Herrera}\ \emph {et~al.}(2016)\citenamefont
  {Torres-Herrera}, \citenamefont {Karp}, \citenamefont {Távora},\ and\
  \citenamefont {Santos}}]{e18100359}%
  \BibitemOpen
  \bibfield  {author} {\bibinfo {author} {\bibfnamefont {E.~J.}\ \bibnamefont
  {Torres-Herrera}}, \bibinfo {author} {\bibfnamefont {J.}~\bibnamefont
  {Karp}}, \bibinfo {author} {\bibfnamefont {M.}~\bibnamefont {Távora}},\ and\
  \bibinfo {author} {\bibfnamefont {L.~F.}\ \bibnamefont {Santos}},\ }\bibfield
   {title} {\bibinfo {title} {Realistic many-body quantum systems vs. full
  random matrices: Static and dynamical properties},\ }\bibfield  {journal}
  {\bibinfo  {journal} {Entropy}\ }\textbf {\bibinfo {volume} {18}},\ \href
  {https://doi.org/10.3390/e18100359} {10.3390/e18100359} (\bibinfo {year}
  {2016})\BibitemShut {NoStop}%
\bibitem [{\citenamefont {Anikeeva}\ \emph {et~al.}(2021)\citenamefont
  {Anikeeva}, \citenamefont {Markovi\ifmmode~\acute{c}\else \'{c}\fi{}},
  \citenamefont {Borish}, \citenamefont {Hines}, \citenamefont {Rajagopal},
  \citenamefont {Cooper}, \citenamefont {Periwal}, \citenamefont
  {Safavi-Naeini}, \citenamefont {Davis},\ and\ \citenamefont
  {Schleier-Smith}}]{PRXQuantum.2.020319}%
  \BibitemOpen
  \bibfield  {author} {\bibinfo {author} {\bibfnamefont {G.}~\bibnamefont
  {Anikeeva}}, \bibinfo {author} {\bibfnamefont {O.}~\bibnamefont
  {Markovi\ifmmode~\acute{c}\else \'{c}\fi{}}}, \bibinfo {author}
  {\bibfnamefont {V.}~\bibnamefont {Borish}}, \bibinfo {author} {\bibfnamefont
  {J.~A.}\ \bibnamefont {Hines}}, \bibinfo {author} {\bibfnamefont {S.~V.}\
  \bibnamefont {Rajagopal}}, \bibinfo {author} {\bibfnamefont {E.~S.}\
  \bibnamefont {Cooper}}, \bibinfo {author} {\bibfnamefont {A.}~\bibnamefont
  {Periwal}}, \bibinfo {author} {\bibfnamefont {A.}~\bibnamefont
  {Safavi-Naeini}}, \bibinfo {author} {\bibfnamefont {E.~J.}\ \bibnamefont
  {Davis}},\ and\ \bibinfo {author} {\bibfnamefont {M.}~\bibnamefont
  {Schleier-Smith}},\ }\bibfield  {title} {\bibinfo {title} {Number
  partitioning with grover's algorithm in central spin systems},\ }\href
  {https://doi.org/10.1103/PRXQuantum.2.020319} {\bibfield  {journal} {\bibinfo
   {journal} {PRX Quantum}\ }\textbf {\bibinfo {volume} {2}},\ \bibinfo {pages}
  {020319} (\bibinfo {year} {2021})}\BibitemShut {NoStop}%
\bibitem [{\citenamefont {Timms}\ \emph {et~al.}(2021)\citenamefont {Timms},
  \citenamefont {Sieberer},\ and\ \citenamefont
  {Kolodrubetz}}]{PhysRevLett.127.270601}%
  \BibitemOpen
  \bibfield  {author} {\bibinfo {author} {\bibfnamefont {C.~I.}\ \bibnamefont
  {Timms}}, \bibinfo {author} {\bibfnamefont {L.~M.}\ \bibnamefont
  {Sieberer}},\ and\ \bibinfo {author} {\bibfnamefont {M.~H.}\ \bibnamefont
  {Kolodrubetz}},\ }\bibfield  {title} {\bibinfo {title} {Quantized floquet
  topology with temporal noise},\ }\href
  {https://doi.org/10.1103/PhysRevLett.127.270601} {\bibfield  {journal}
  {\bibinfo  {journal} {Phys. Rev. Lett.}\ }\textbf {\bibinfo {volume} {127}},\
  \bibinfo {pages} {270601} (\bibinfo {year} {2021})}\BibitemShut {NoStop}%
\bibitem [{\citenamefont {von Keyserlingk}\ \emph {et~al.}(2016)\citenamefont
  {von Keyserlingk}, \citenamefont {Khemani},\ and\ \citenamefont
  {Sondhi}}]{PhysRevB.94.085112}%
  \BibitemOpen
  \bibfield  {author} {\bibinfo {author} {\bibfnamefont {C.~W.}\ \bibnamefont
  {von Keyserlingk}}, \bibinfo {author} {\bibfnamefont {V.}~\bibnamefont
  {Khemani}},\ and\ \bibinfo {author} {\bibfnamefont {S.~L.}\ \bibnamefont
  {Sondhi}},\ }\bibfield  {title} {\bibinfo {title} {Absolute stability and
  spatiotemporal long-range order in floquet systems},\ }\href
  {https://doi.org/10.1103/PhysRevB.94.085112} {\bibfield  {journal} {\bibinfo
  {journal} {Phys. Rev. B}\ }\textbf {\bibinfo {volume} {94}},\ \bibinfo
  {pages} {085112} (\bibinfo {year} {2016})}\BibitemShut {NoStop}%
\bibitem [{\citenamefont {Lazarides}\ and\ \citenamefont
  {Moessner}(2017)}]{PhysRevB.95.195135}%
  \BibitemOpen
  \bibfield  {author} {\bibinfo {author} {\bibfnamefont {A.}~\bibnamefont
  {Lazarides}}\ and\ \bibinfo {author} {\bibfnamefont {R.}~\bibnamefont
  {Moessner}},\ }\bibfield  {title} {\bibinfo {title} {Fate of a discrete time
  crystal in an open system},\ }\href
  {https://doi.org/10.1103/PhysRevB.95.195135} {\bibfield  {journal} {\bibinfo
  {journal} {Phys. Rev. B}\ }\textbf {\bibinfo {volume} {95}},\ \bibinfo
  {pages} {19Dear Editor,} (\bibinfo {year} {2017})}\BibitemShut {NoStop}%
\bibitem [{\citenamefont {Martyn}\ \emph {et~al.}(2021)\citenamefont {Martyn},
  \citenamefont {Rossi}, \citenamefont {Tan},\ and\ \citenamefont
  {Chuang}}]{PRXQuantum.2.040203}%
  \BibitemOpen
  \bibfield  {author} {\bibinfo {author} {\bibfnamefont {J.~M.}\ \bibnamefont
  {Martyn}}, \bibinfo {author} {\bibfnamefont {Z.~M.}\ \bibnamefont {Rossi}},
  \bibinfo {author} {\bibfnamefont {A.~K.}\ \bibnamefont {Tan}},\ and\ \bibinfo
  {author} {\bibfnamefont {I.~L.}\ \bibnamefont {Chuang}},\ }\bibfield  {title}
  {\bibinfo {title} {Grand unification of quantum algorithms},\ }\href
  {https://doi.org/10.1103/PRXQuantum.2.040203} {\bibfield  {journal} {\bibinfo
   {journal} {PRX Quantum}\ }\textbf {\bibinfo {volume} {2}},\ \bibinfo {pages}
  {040203} (\bibinfo {year} {2021})}\BibitemShut {NoStop}%
\end{thebibliography}%
\end{document}